\documentclass[12pt]{article}
\usepackage{a4wide}
\usepackage{amssymb}
\usepackage{amsmath}
\usepackage{graphicx}
\usepackage{mathtools}
\usepackage{caption}
\usepackage{float}

\begin{document}
{\renewcommand{\thefootnote}{\fnsymbol{footnote}}
\begin{center}
{\LARGE  Canonical Tunneling Time in Ionization Experiments}\\
\vspace{1.5em}
Bekir Bayta\c{s}\footnote{e-mail address: {\tt bub188@psu.edu}},
Martin Bojowald\footnote{e-mail address: {\tt bojowald@gravity.psu.edu}}
and Sean Crowe\footnote{e-mail address: {\tt stc151@psu.edu}}
\\
\vspace{0.5em}
Department of Physics,
The Pennsylvania State
University,\\
104 Davey Lab, University Park, PA 16802, USA\\
\vspace{1.5em}
\end{center}
}

\setcounter{footnote}{0}

\begin{abstract}
  Canonical semiclassical methods can be used to develop an intuitive
  definition of tunneling time through potential barriers. An application to
  atomic ionization is given here, considering both static and time-dependent
  electric fields. The results allow one to analyze different theoretical
  constructions proposed recently to evaluate ionization experiments based on
  attoclocks. They also suggest new proposals of determining tunneling times,
  for instance through the behavior of fluctuations.
\end{abstract}

\section{Introduction}

Detailed observations of atom ionization have recently become possible with
attoclock experiments \cite{TunnelHelium,TunnelExperiment, TunnelingHydrogen},
suggesting comparisons with various predictions of tunneling times. The
theoretical side of the question, however, remains largely open: Different
proposals of how to define tunneling times have been made through almost nine
decades, yielding widely diverging predictions and physical interpretations
\cite{TunnelingReview,BarrierTime}. Even the extraction of tunneling times
from experiments has been performed in different ways \cite{TunnelWeakMeas,
  TunnelingResolution,
  Interpreting,TunnelingCoord,TrajectoryFree,TimeMomentum}, and the original
conclusion of a non-zero result has been challenged
\cite{BackProp,BackProp2,BackProp3}. The situation therefore remains far from
being clarified, and a continuing analysis of fundamental aspects of tunneling
is important.

A recent approach to understand the tunneling dynamics in this context is the
application of Bohmian quantum mechanics \cite{BohmI,BohmII}, in which the
prominent role played by trajectories provides a more direct handle on
tunneling times \cite{Bohmian}.  However, through initial conditions, the
ensemble of trajectories remains subject to statistical fluctuations. An
alternative trajectory approach, which we will develop in this paper, is to
consider, in an extension of Ehrenfest's theorem, the evolution of expectation
values and fluctuations, possibly together with higher-order moments of a
state. By including moments of a probability distribution, such an approach
remains statistical in order to capture quantum properties, but it provides a
unique trajectory starting with the expectation values and fluctuations of a
given initial state. The ensemble of trajectories used in Bohmian quantum
mechanics is replaced by a single trajectory in an extended phase space,
enlarged by fluctuations and higher moments as non-classical dimensions.

In the context of tunneling, a semiclassical version of this proposal has been
used occasionally in quantum chemistry \cite{GaussianDyn,QHDTunneling}, which
we extend here to higher orders and apply to models of atom ionization.
Unlike Bohmian quantum mechanics, these methods present an approximation to
standard quantum mechanics, rather than a new formulation. Nevertheless, since
they lead to a single trajectory rather than a statistical ensemble of
trajectories, they provide a crucial advantage which, we hope, can help to
clarify the question of tunneling times in atom ionization.

In \cite{Bohmian}, it has been shown that a trajectory approach based on
Bohmian quantum mechanics reliably shows non-zero tunneling times in atomic
models of ionization. There is therefore a tension with recent evaluations of
ionization experiments which give the impression of zero tunneling delays
\cite{BackProp2}. The latter results are based on a definition of the
tunneling exit time through classical back-propagation \cite{BackProp}: Since
the energy of a tunneling electron in a time-dependent electric field is not
conserved and usually unkonwn in experiments, it is difficult to apply the
intuitive definition of the tunneling exit as the time when the electron's
energy equals the classical potential. As an alternative, classical
back-propagation evolves the final state of a measured electron back toward
the atom using classical equations of motion, and defines the tunneling exit
as the time when the momentum in the direction of the electric field is zero,
taking the point closest to the atom in the event that this condition may be
realized multiple times. As already noted in \cite{Bohmian}, this condition is
conceptually problematic because it uses classical physics near a turning
point, where the equations governing a classically back-propagated trajectory
are usually expected to break down. We will use our single-trajectory approach
to compare a quantum trajectory with a classical back-propagated one.

In addition, our analysis will allow us to derive further properties of the
tunneling process.  In order to obtain a single trajectory describing an
evolving quantum state, we write evolution of a quantum state in terms of a
classical-type system with quantum corrections, in which the expectation
values of position and momentum are coupled to fluctuations. The coupling
terms, quite generally, lower the classical barrier such that the
classical-type system can move ``around'' it in an extended phase space
with a real-valued velocity. This detour has a certain duration,
depending on initial conditions, and provides a natural definition of
tunneling time.

It turns out that several new ingredients are necessary compared with existing
treatments in quantum chemistry. For instance, semiclassical states are not
always sufficient for a full description of tunneling. This fact is not
surprising because, intuitively, a tunneling wave splits up into two wave
packets separated by the barrier width. Deep tunneling then implies states
with large fluctuations, even if each wave packet remains sharp and perhaps
nearly Gaussian. Moreover, fluctuation terms do not always lower the barrier
enough to make tunneling possible at all energies for which quantum tunneling
occurs. In \cite{QHDTunneling}, the classical-type system used for tunneling
has been extended to moments of up to fourth order, with a clear improvement
of predicted tunneling times closer to what follows from wave-function
evolution. However, the extension was done mainly at a numerical level, which
does not provide much intuition about the tunneling process in a given
potential. To second order, by contrast, an effective potential was used in
\cite{GaussianDyn,QHDTunneling} which shows how the classical barrier can be
lowered by quantum fluctuations. One of our main new ingredients is an
extension of such effective potentials to higher orders.

In Sec.~\ref{sec:Quantum dynamics} we describe quantum dynamics using
canonical semiclassical methods and present a new effective potential that
includes effects from higher-order moments. In Sec.~\ref{sec:Models}, we
introduce various models of atom ionization in which our methods can be
applied, and discuss specific results focusing on tests of tunneling
conditions and the definition of tunneling times.

\section{Quantum dynamics by canonical effective methods} 
\label{sec:Quantum dynamics}

Using canonical effective methods \cite{EffAc,Karpacz}, we describe the
dynamics of a quantum state by coupled ordinary differential equations for the
expectation values $x=\langle\hat{x}\rangle$ and $p=\langle\hat{p}\rangle$
coupled to central, Weyl-ordered moments
\begin{equation}
 \Delta(x^ap^b)= \langle(\hat{x}-x)^a(\hat{p}-p)^b\rangle_{\rm Weyl}\,.
\end{equation}
(In this notation, the usual fluctuations are written as $\Delta(x^2)=(\Delta
x)^2$ and $\Delta(p^2)=(\Delta p)^2$, while $\Delta(xp)$ is the covariance.)

The Hamiltonian operator $H(\hat{x},\hat{p})$ implies the quantum Hamiltonian
\begin{eqnarray} \label{HQ}
 H_{{\rm Q}}
&=& \langle H(\hat{x} + (\hat{x} - x), \hat{p} + (\hat{p} -
p))\rangle \nonumber\\ 
&=& H(x,p)+ \sum_{n=2}^{\infty} \sum_{a=0}^n
\left(\begin{array}{c}n\\a\end{array}\right) 
 \frac{\partial^n H(x,p)}{\partial x^a\partial p^{n-a}}
\Delta(x^ap^{n-a})
\end{eqnarray}
with the classical Hamiltonian $H(x,p)$. Hamiltonian equations for moments are
generated using the Poisson bracket 
\begin{equation}
 \{\langle\hat{A}\rangle,\langle\hat{B}\rangle\}=
 \frac{\langle[\hat{A},\hat{B}]\rangle}{i\hbar}\,,
\end{equation}
derived from the commutator and extended to moments by using linearity and the
Leibniz rule.

Unfortunately, the Poisson brackets between moments are rather complicated at
higher orders, and they are not canonical. For instance,
\begin{equation}
 \{\Delta(x^2),\Delta(xp)\}=2\Delta(x^2)\quad,\quad 
\{\Delta(x^2),\Delta(p^2)\}=4\Delta(xp)\quad,\quad
\{\Delta(xp),\Delta(p^2)\}=2\Delta(p^2)\,,
\end{equation}
corresponding to the Lie algebra ${\rm sp}(2,{\mathbb R})$, but those of
higher moments are in general non-linear. For these second-order moments,
canonical variables were introduced in \cite{GaussianDyn,QHDTunneling}:
\begin{equation}
 s=\sqrt{\Delta(x^2)}\quad,\quad p_s= \frac{\Delta(xp)}{\sqrt{\Delta(x^2)}}
\end{equation}
together with a third variable, $U=\Delta(x^2)\Delta(p^2)-\Delta(xp)^2$, which
has zero Poisson brackets with $s$ and $p_s$. Inverting these relationships,
we write the second-order moments
\begin{equation} \label{Deltas}
 \Delta(x^2)=s^2\quad,\quad \Delta(xp)=sp_s\quad,\quad
 \Delta(p^2)=p_s^2+\frac{U}{s^2}
\end{equation}
in terms of canonical variables $(s,p_s)$ and a conserved quantity $U$.  To
second order, the quantum Hamiltonian can then be expressed as
\begin{eqnarray} \label{H}
 \langle\hat{H}\rangle &=& \frac{\langle\hat{p}^2\rangle}{2m}+ \langle
 V(\hat{x})\rangle\nonumber\\
&\approx& \frac{\langle\hat{p}\rangle^2}{2m} + \frac{(\Delta p)^2}{2m}
+V(\langle\hat{x}\rangle)+ \frac{1}{2} V''(\langle\hat{x}\rangle) (\Delta
x)^2 = \frac{p^2+p_s^2}{2m}+V_{\rm eff}(x,s)
\end{eqnarray}
with the effective potential
\begin{equation} \label{VEffSemi}
 V_{\rm eff}(x,s)=  V(x) +\frac{U}{2ms^2}+
\frac{1}{2}V''(x)  s^2\,.
\end{equation}

An extension to higher orders turns out to be more involved, but it can be
accomplished with the new methods developed in \cite{Bosonize}. The canonical
form of higher-order moments then gives useful higher-order effective
potentials, and it suggests closure conditions, in the sense of
\cite{Closure}, that can be used to turn the infinite set of moments into
finite approximations. 

We introduce closure conditions based on the following properties of higher
moments which we have confirmed for up to fourth order \cite{EffPotRealize}:
the second-order variable $s$ also contributes to an $n$-th order moment, in
the form $\langle (\hat{x}-\langle \hat{x}\rangle)^n\rangle \approx s^n$, in
addition to terms that depend on new degrees of freedom.  Moments of odd and
even order, respectively, often behave rather differently from each other. For
instance, a Gaussian has zero odd-order moments, a property which extends to
generic states that evolve adiabatically in symmetric potentials
\cite{EffAc}. This difference is reflected in mathematical properties of the
canonical variables. At third order, for instance, there are three canonical
coordinates, $s_1$, $s_2$ and $s_3$, such that
$\langle(\hat{x}-\langle\hat{x}\rangle)^3\rangle\propto
s_1^3+s_2^3+s_3^3$. The constant of proportionality has zero Poisson brackets
with the canonical variables but is state dependent.  As an approximation, we
set this constant equal to zero, reducing the number of degrees of freedom.
If we assume this behavior also for orders greater than four, we can complete
the Taylor expansion in (\ref{HQ}) and derive the all-orders effective
potential
\begin{eqnarray} \label{VEff}
 V_{\rm eff}(x,s) &=& \frac{U}{2ms^2} + V(x) + \sum_{n=1}^{\infty}
  \frac{1}{2 n!} \frac{{\rm d}^{2 n}(V(x))}{{\rm d}x^{2 n}} s^{2 n}\nonumber\\
 &=&  \frac{U}{2ms^2} +\frac{1}{2}\left( V(x+s) + V(x-s)\right)  \,.
\end{eqnarray}
Heuristically, therefore, the particle does not follow a potential local in
$x$, but rather is feeling around itself at a distance $s$. This distance
increases as the wave function spreads out.

We have moved beyond the semi-classical approximation by replacing a strict
truncation with a specific behavior of the moments. This extension is crucial
for our purposes because tunneling states or the ground states of an electron
in most atoms are not semi-classical. A semi-classical approximation should
then not be expected to give accurate results in situations where the
tunneling times are very long, or the electron spends a fair amount of time in
states close to the ground state.

\section{Effective theory of tunneling ionization}
\label{sec:Models}

In order to test various aspects that have been found to be relevant for
tunneling times in ionization experiments, we discuss properties and results
of different models.  An application to tunneling ionization requires an
extension of (\ref{VEff}) to three dimensions. The main question is then how
to deal with cross-correlations between different coordinates, which
significantly enlarge the phase space. Motivated by the intuition that a
tunneling wave packet should split up predominantly in the direction of the
force that lowers the confining potential of a bound state, we assume that the
main moments to be considered are the two fluctuations (position and momentum)
in the direction of the force. These moments then play the role of reaction
coordinates \cite{TransitionState}, which reduce a large parameter space to a
few significant variables.

The relationship to the direction of the force implies a crucial
difference between the treatment of a constant force and time-dependent,
rotating forces as used in attoclock experiments. We first deal with examples
subject to a constant force in order to illustrate the tunneling process with
our new methods, and then show how time-dependent forces alter the
conclusions.

\subsection{Coulomb potential in a static electromagnetic field}

As usual, we can treat tunneling ionization as a a single electron moving in
an effective potential with two contributions: a spherically symmetric term
for interactions with the nucleus and the remaining electrons, and a linear
potential in the direction of the electric field.  Assuming that correlations
between the independent coordinates can be ignored, an approximation that can
be expected to be valid during most of the tunneling process which affects
mainly one of the coordinates, the all-orders effective potential (\ref{VEff})
for the 3-dimensional Coulomb interaction and the electric field strength $F$
is then
\begin{equation} \label{VEff3D}
V_{\rm eff}(x_i,s_j)= \sum \limits_{i=1}^3 \frac{U}{2 s_i^2} +
\frac{1}{8} \sum \limits_{\{n_i=0,1\}} V\left(x_i +(-1)^{n_i} s_i\right)  \,, 
\end{equation}
where 
\begin{equation} \label{VClass}
V(\vec{x}) = - \frac{1}{|\vec{x}|} - \vec{x} \cdot \vec{F} - \frac{\alpha_I
  \vec{F} \cdot \vec{x}}{|\vec{x}|^3} 
\end{equation}
is the classical potential and $\alpha_I$ is the static polarizability of the
ion. (We set $\vec{x}=(x,y,z)$ and use atomic units $\hbar=e=m_e=k_e=1$
throughout the paper.)
     
Evolution in the effective potential requires initial values of $x_i$, $s_i$,
$p_{i}$ and $p_{s_i}$. Since these describe expectation values and
fluctuations, they could in principle be determined from an initial atomic
state. However, it is more useful to minimize the energy in the field-free
($\vec{F}=0$) effective potential (\ref{VEff3D}), in order to fix these
initial values within our approximation. That is, to get initial values for
the canonical variables we minimize $\frac{1}{2}\sum (p_i^2+p_{s_i}^2) +
V_{\rm eff}(\vec{x},\vec{s})$ in the absence of the electric field. We find
\begin{eqnarray} \label{InitialValues}
s_i^0 = \frac{3 \sqrt{3}}{4} \quad \mathrm{and} \quad p_i^0=p_{s_i}^0=x_i^0=0
\end{eqnarray}
for $i=1,2,3$.  These values, taken as initial conditions for tunneling with a
non-zero field, result in a ionization potential of $I_p=-2/9$ which in our
model corresponds the ground-state energy $E_{\rm ground}$ in the absence of
the electric field.

  \begin{figure}[htbp]
\begin{center}
      \includegraphics[scale=1]{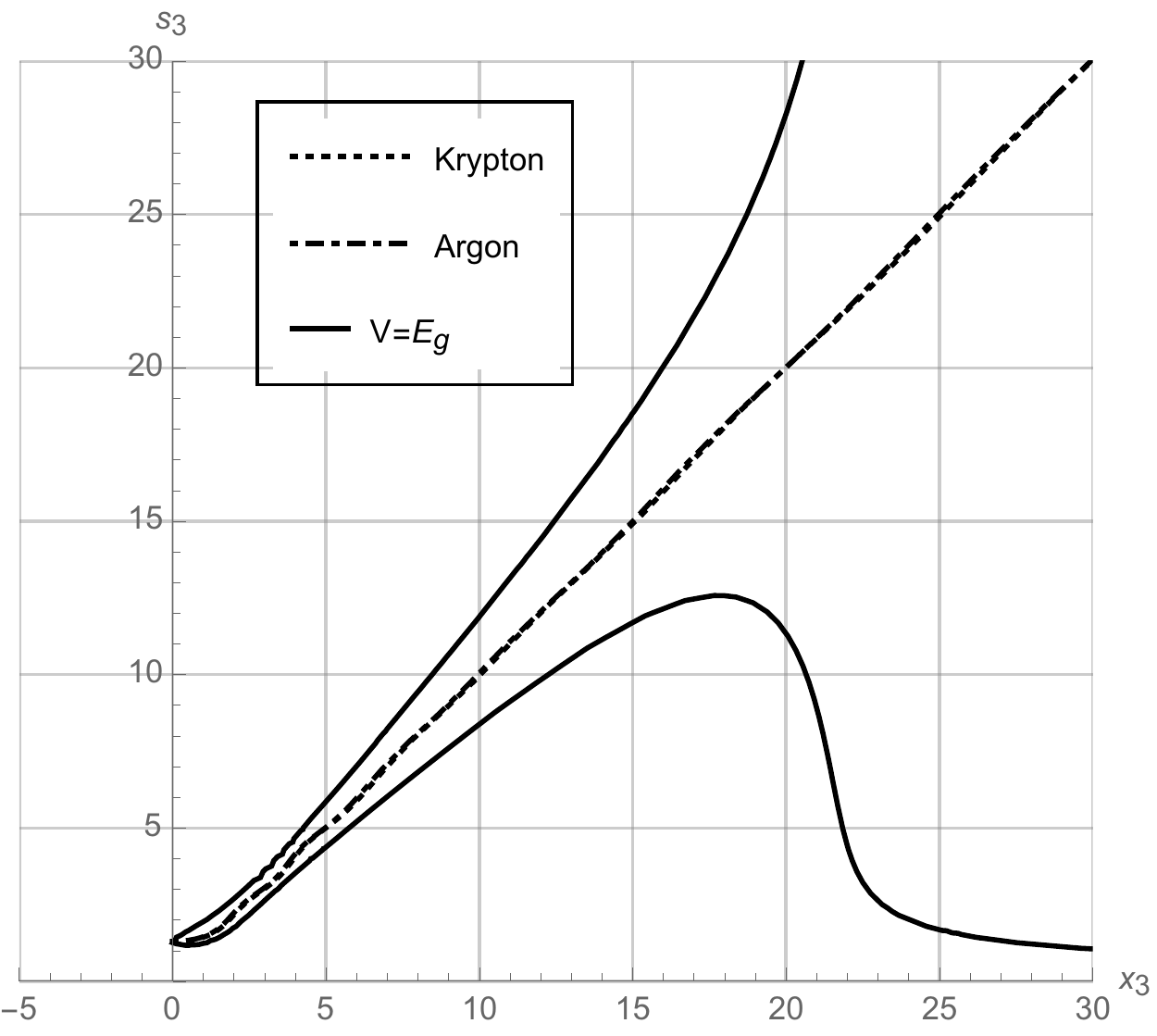}
      \caption{A contour plot of the effective potential for both Argon
        ($\alpha_I =7$) and Krypton ($\alpha_I =11$). The solid curve is the
        equipotential line, $V_{\rm eff}=E_{\rm ground}=-2/9$ for the
        approximate ground-state energy corresponding to
        (\ref{InitialValues}). It shows the location of the classical barrier
        in the presence of a field $F = 0.015$ (a laser intensity of $I \sim
        0.8\cdot 10^{14}{\rm W}/{\rm cm}^2$). The path of the electron is
        shown here by the (almost overlapping) dashed lines for Argon and
        Krypton. The electron escaping from either
        atom has to travel along an actual tunnel, formed by the
        equi-potential line in phase space. \label{ContourPlot}}
\end{center}
\end{figure}     

We choose our coordinate system such that the $x_3$-axis points in the
direction of the force. Figure~\ref{ContourPlot} shows the ground-state
equipotential line of (\ref{VEff3D}) in the $x_3-s_3$ plane for both Argon
($\alpha_I=7$) and Krypton ($\alpha_I=11$), as well as the behavior of the
fluctuation parameter $s_3$ with respect to the direction along $x_3$. When
the field strength is small enough, the equipotential line of the ground state
literally forms a tunnel that the electron has to follow in order to
escape. The tunneling time is related to the amount of time spent in this
tunnel. At this point, we can see the importance of our extension beyond
semiclassical effective potentials. The quadratic $s$-term in (\ref{VEffSemi})
reduces the classical barrier monotonically in the $s$-direction, giving us a
steep slope instead of a tunnel. Numerical solutions in such a potential show
that the resulting tunneling times would be too large because trajectories get
dragged into the $s$-direction with little movement in the $x$-direction. The
tunnel in our all-orders potential, by contrast, guides the trajectories such
that they still move substantially in the $x_3$-direction. Corresponding
tunneling times are significantly shorter.

\begin{figure}[htbp]
\begin{center}
\includegraphics[scale = 0.75]{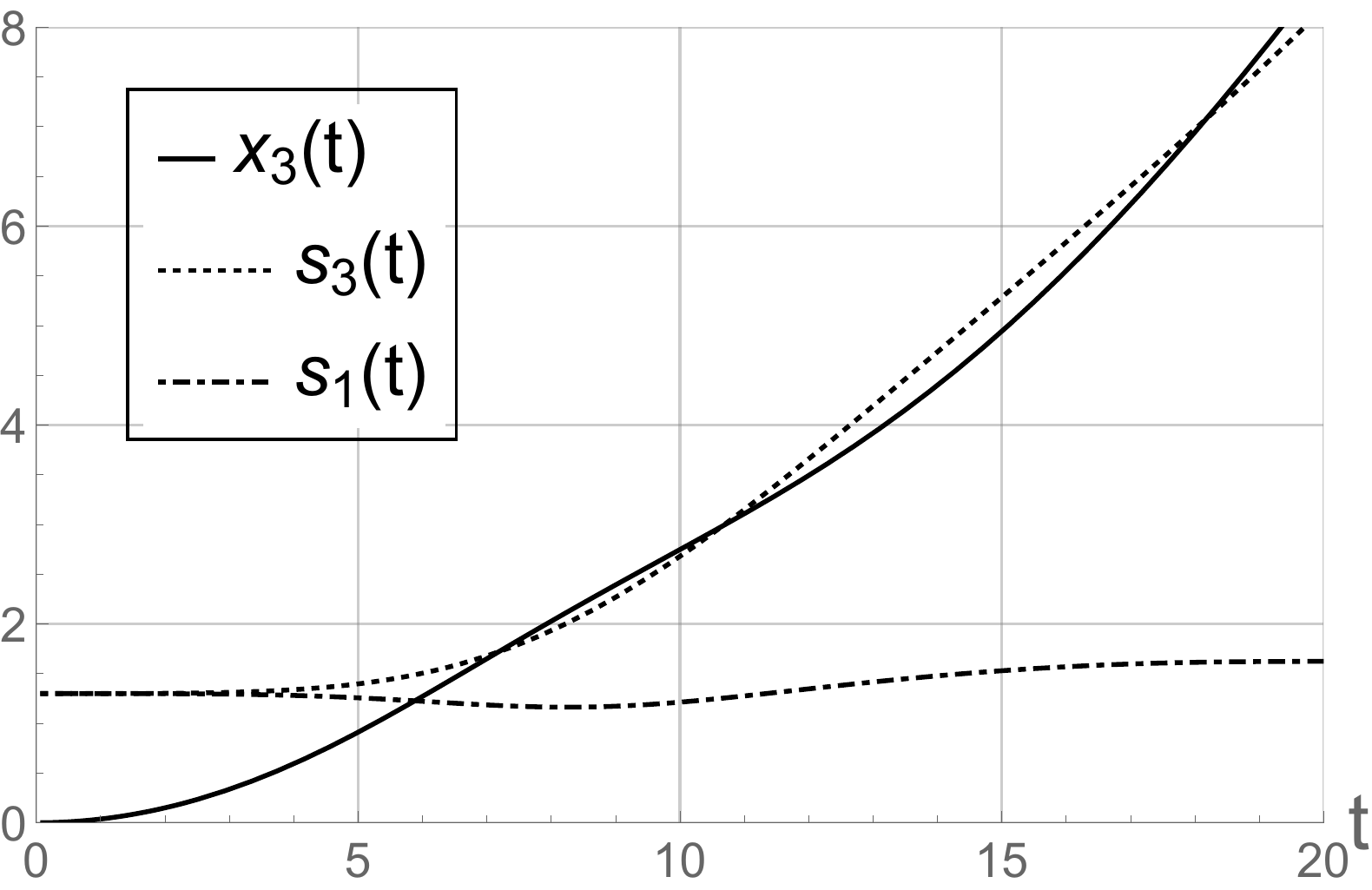}
\caption{Trajectories of the tunneling coordinate $x_3$, its
  fluctuation $s_3$ and the fluctuation $s_1$ for Argon. The behavior for
  Krypton is qualitatively similar. \label{Evolution}}
\end{center}
\end{figure}

\begin{figure}[htbp]
\begin{center}
\includegraphics[scale = 1.0]{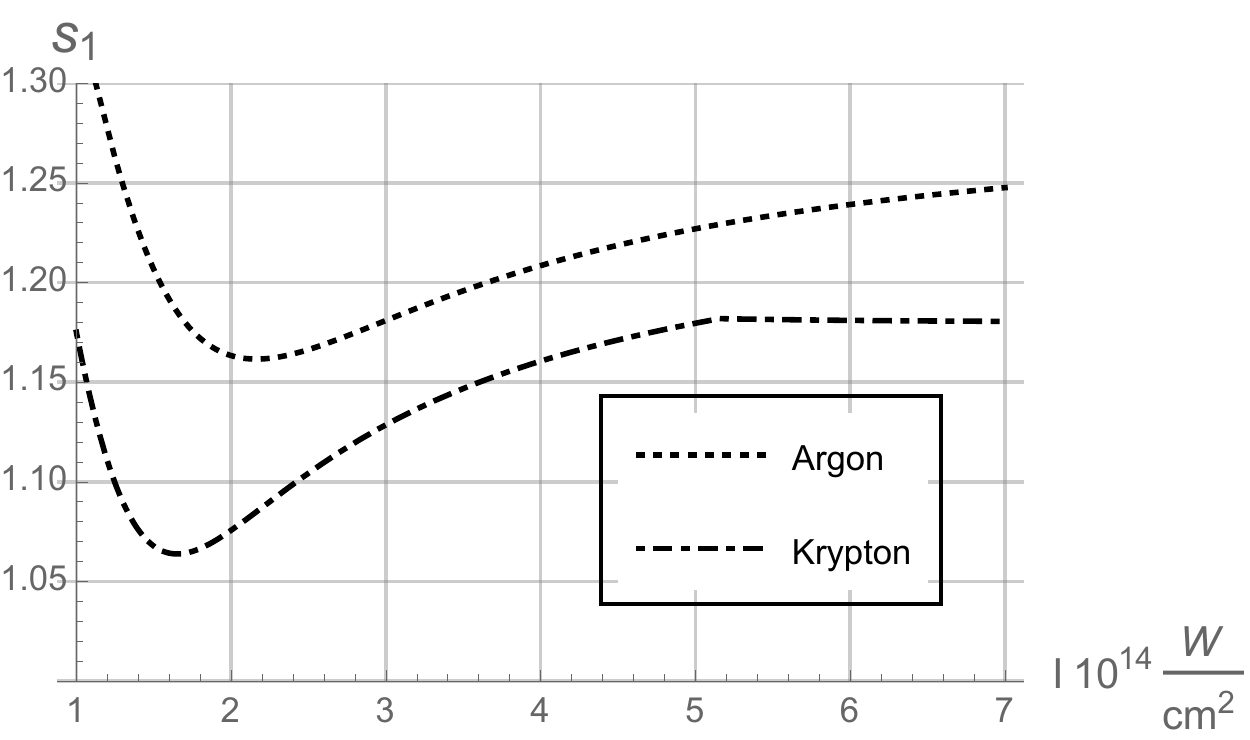}
\caption{The transverse exit fluctuation $s_1$ over the observable
  range of laser intensities for Argon and Krypton. \label{Fluctuations}}
\end{center}
\end{figure}

Our dynamical system contains not only expectation values but also the
fluctuation variables $s_i$ and $p_{s_i}$, related to $\Delta x_i$ and $\Delta
p_i$ as in (\ref{Deltas}). As shown in Fig.~\ref{Evolution}, our effective
evolution is self-consistent in the sense that it is indeed only the
fluctuation $s_3$ in the direction of the force (our reaction coordinate) that
increases significantly, while $s_1$ and $s_2$ remain nearly constant.
Nevertheless, the behavior of the transversal position fluctations, shown in
Fig.~\ref{Fluctuations} for the example of $s_1$ at the tunneling exit, is
also of interest: There is a local minimum with a value less than the
ground-state fluctuation (\ref{InitialValues}). At higher intensities, the
fluctuations level off because in a strong field they do not have much time to
change. Moreover, these fluctuations depend more strongly on the element used
compared to the trajectories in Fig.~\ref{ContourPlot} for variables in the
direction of the force, or the tunneling time to which we turn now.

\begin{figure}[htbp]
\begin{center}
\includegraphics[width=14cm]{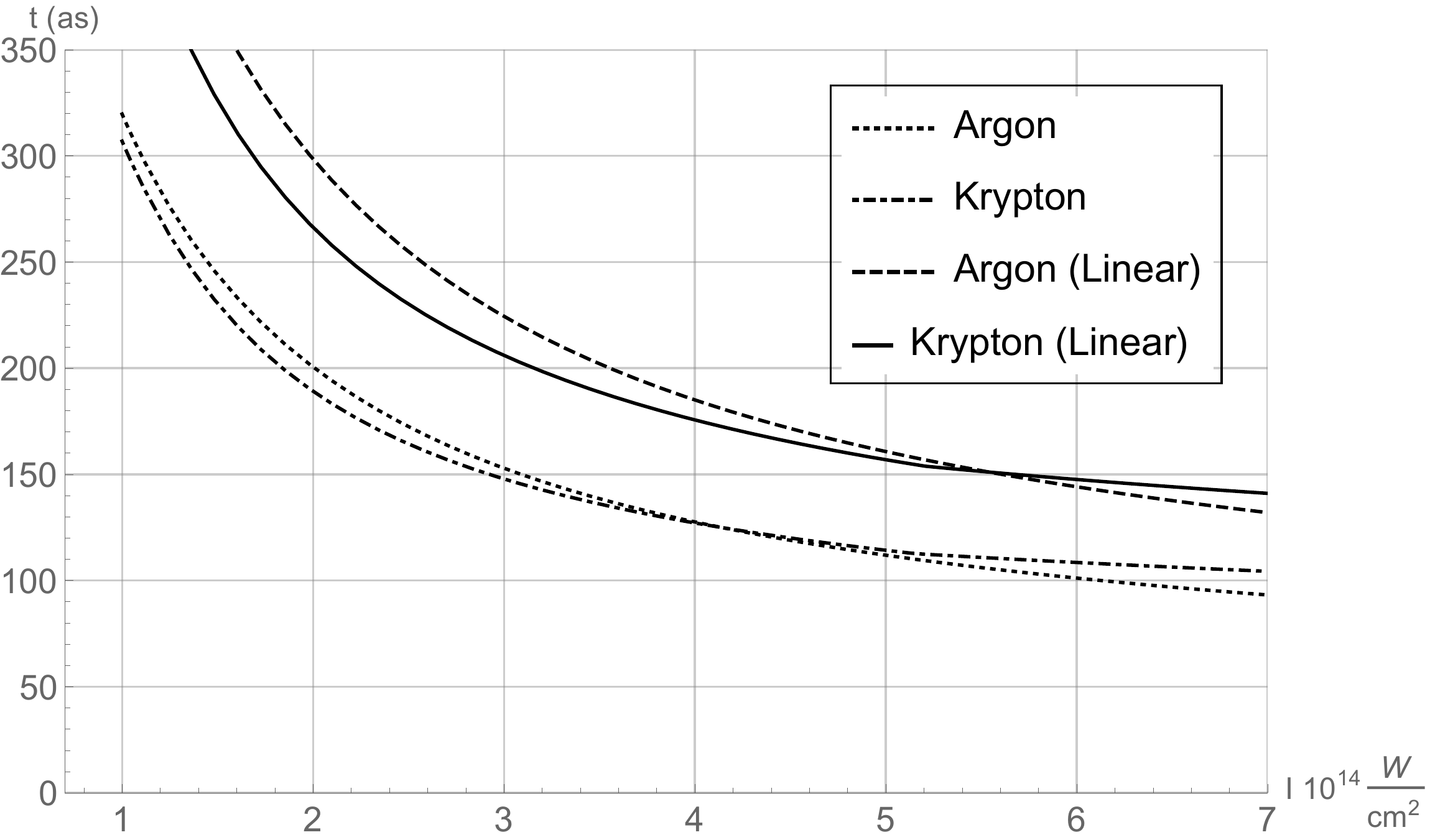}
\caption{Tunneling times for Argon and Krypton. The dashed (Argon) and solid
  (Krypton) lines correspond to the approximation (\ref{tauapprox}) with $s_3
  \approx x_3$.  The range of the laser intensity is obtained by scaling the
  electric field $I =\frac{1}{2}c \epsilon_0 F^2$. Time variables are scaled
  to atto-seconds from atomic units. \label{tunneling-time}}
\end{center}
\end{figure}

Using the all-orders potential in a static field, we estimate the tunneling
time in Argon and Krypton as a function of the laser intensity. The tunneling
time is determined by how long the particle travels from one turning point to
another in a state parameterized by $x_i$ and $s_i$. The tunneling times for
both Argon and Krypton in the range of laser intensities used in
\cite{TunnelExperiment}, are shown in Fig.~\ref{tunneling-time}. We see
tunneling at all relevant scales, and qualitative agreement with the
calculations from Wigner formalism used in \cite{TunnelExperiment}.

Traditionally, proposed tunneling times have often been expressed as integral
formulas, motivated by the WKB approximation. Our effective potential can be
used to derive a new version if we eliminate some of the basic variables in
further approximations. As suggested by Figs.~\ref{ContourPlot} and
\ref{Evolution}, we may assume that $s_3 \approx x_3$ inside the barrier.  The
tunneling time can then be written as
\begin{equation} \label{tauapprox}
\tau \approx  \int^{x_3^*}_{0} \frac{{\rm d}x_3}{p_3} \approx  \int^{x_3^*}_{0}
\frac{{\rm d}x_3}{\sqrt{-E_{\rm ground} - V_{\rm{eff}}(x_i,\tilde{s}_i)}} \, , 
\end{equation}
where $\tilde{s}_3 = x_3$ and $x_3^*$ is the tunneling exit position.  The
values of $x_1$ and $x_2$ are assumed zero, while $\tilde{s}_1$ and
$\tilde{s}_2$ retain their ground-state values. The qualitative behavior of
the tunneling time in Fig.~\ref{tunneling-time} under this approximation is
not too far from the results of our full computation.

\begin{figure}[htbp]
\begin{center}
\includegraphics[width=14cm]{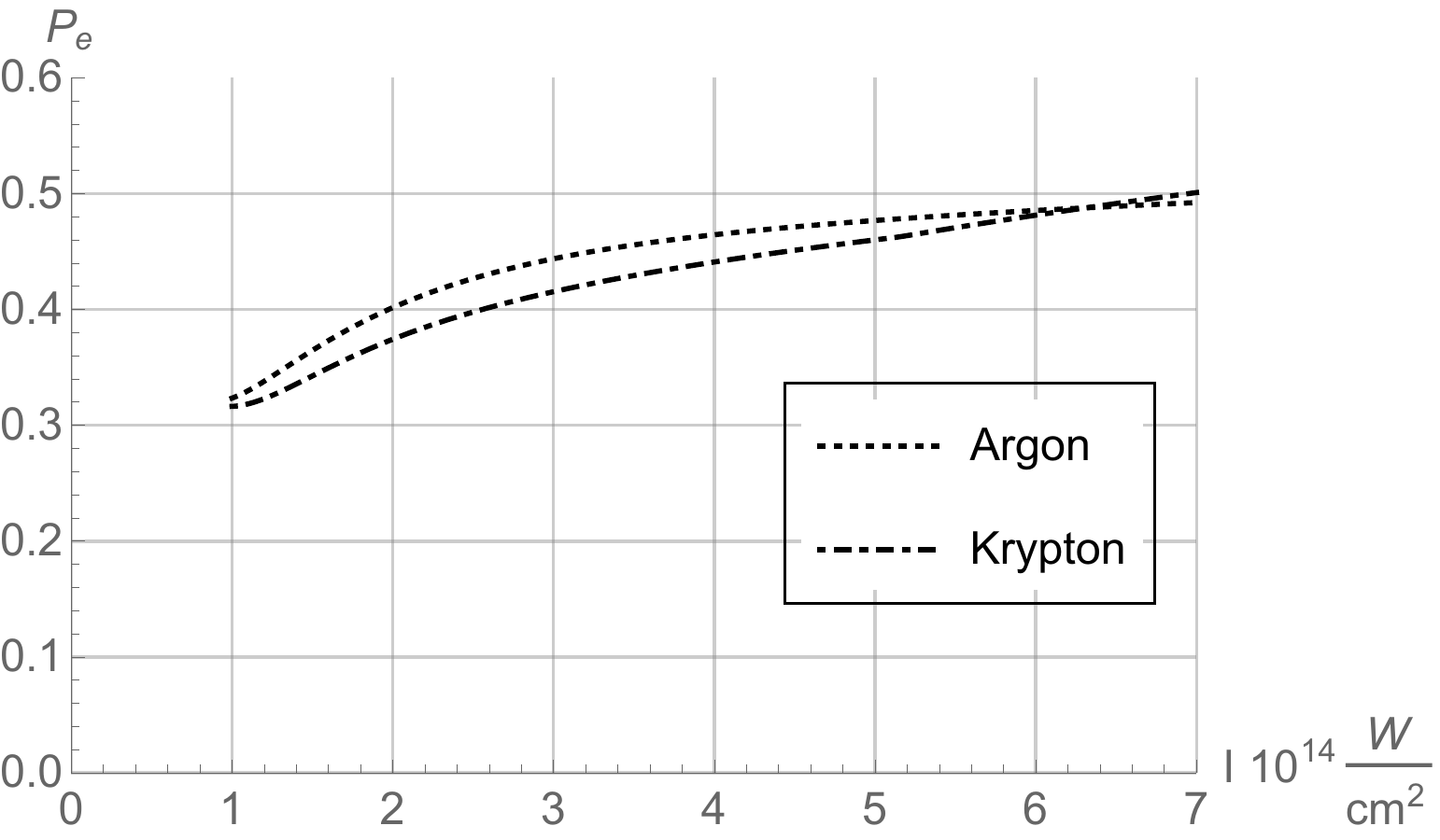}
\caption{Exit momenta for the electron as a function of the laser
intensity. They have the same qualitative behavior as in
\cite{TunnelExperiment} with an agreement of order of magnitude.
\label{momentum} }
\end{center}
\end{figure}

Our method also yields the momentum $p_3$ at the tunnel exit, shown in
Fig.~\ref{momentum}. The longitudinal momentum is non-zero because the
electron exits the tunnel with momentum in the direction of the force: As
shown in Fig.~\ref{ContourPlot}, in the effective potential, the classical
turning point is replaced by an actual tunnel exit. Our effective potential
therefore presents a self-contained model in which several observational
features are qualitatively reproduced, without any free parameters beyond the
coefficients used to define the classical potential. However, it requires an
extension to time-dependent forces modelling laser fields.

\subsection{Time-dependent, circularly polarized electric fields}

If the direction of the force is not constant, tunneling should affect the
moments of more than one degree of freedom. If the force is rotating at
constant angular velocity $\omega$, we can nevertheless find suitable reaction
coordinates by transforming to a frame co-rotating with the force. It is
sufficient to start with a two-dimensional system in the plane in which the
force is rotating. For instance, the example used in \cite{BackProp2} is a
two-dimensional, time-dependent vector potential
\begin{equation}
 \vec{A}(t) = \frac{A_0}{\sqrt{1+\epsilon^2}} \cos^4(\omega t/2N)
 \left(\begin{array}{c} \cos(\omega t) \\ \epsilon \sin(\omega
     t)\end{array}\right)
\end{equation}
for $N$ cycles of frequency $\omega$, with ellipticity $\epsilon$. The
corresponding electric field is
\begin{equation}
 \vec{E}= -\frac{{\rm d}\vec{A}}{{\rm d}t} =
 \frac{A_0\omega}{\sqrt{1+\epsilon^2}} \cos^4(\omega t/2N)
 \left(\begin{array}{c} \sin(\omega t)+ \frac{2}{N} \tan(\omega t/2N)
     \cos(\omega t) \\ \epsilon\left(-\cos(\omega t)+ \frac{2}{N} \tan(\omega
       t/2N) \sin(\omega t)\right)\end{array}\right)\,.
\end{equation}

Specialized to two cycles, $N=2$, and circular polarization, $\epsilon=1$,
also as in \cite{BackProp2}, we have
\begin{equation}
 \vec{E} = \frac{A_0\omega}{\sqrt{2}} \cos^3(\omega t/4)
 \left(\begin{array}{c} \sin(5\omega t/4)\\ \cos(5\omega
     t/4)\end{array}\right)= \frac{A_0\omega}{\sqrt{2}} \cos^3(\omega t/4)
 S\left(\begin{array}{c}1\\0\end{array}\right)
\end{equation}
with the orthogonal matrix 
\begin{equation}
 S = \left(\begin{array}{cc} \sin(5\omega t/4)&-\cos(5\omega t/4)\\
     \cos(5\omega t/4)&\sin(5\omega t/4)\end{array}\right)\,.
\end{equation}

In terms of the electric field, we can write the Hamiltonian for a negatively
charged particle as
\begin{equation}
 H = \frac{1}{2}\vec{p}^2+ \vec{r}\cdot\vec{E} +V(r)\,.
\end{equation}
In co-rotating coordinates
\begin{equation}
 \vec{R}=S^{-1}\vec{r} \quad,\quad \vec{P} = S^{-1}\vec{p}
\end{equation}
we have
\begin{equation}
 H = \frac{1}{2}\vec{P}^2+ \vec{R}\cdot\vec{E}_0+ V(R)+\frac{5 \omega}{4}\left(P_1 R_2-P_2 R_1\right)
\end{equation}
with an electric field
\begin{equation} \label{Ecos}
 \vec{E}_0= S^{-1}\vec{F}= \frac{A_0\omega}{\sqrt{2}} \cos^3(\omega t/4)
 \left(\begin{array}{c}1\\0\end{array}\right) \, ,
\end{equation}
which is not constant but points in a fixed direction. The fluctuations in
this direction are our reaction coordinates. 

The transformation to a co-rotating frame shows that the two-dimensional
nature of tunneling in circularly polarized electric fields is not essential,
but it turns out that the non-static behavior of the field amplitude is
important. This behavior can be studied by Bohmian quantum mechanics in
one-dimensional models \cite{Bohmian}, or by our effective potentials as we
will do in the rest of this paper.

For our methods, in the one dimensional case, it is of advantage to have a
smooth potential which is finite everywhere. Instead of the Coulomb potential
or the truncated version of \cite{Bohmian}, we therefore consider a
one-dimensional model for a Gaussian potential well in a time-dependent
electric field:
\begin{equation}
\label{eq:Gaussian}
V(x,t) = - \, \frac{e^{-x^2}}{2}  + x \, F(t)  \, .
\end{equation}
The potential depth is chosen so that the ground state energy agrees with
$E_{\rm ground}$. As the time-dependent electric field, we choose, as in
\cite{Bohmian},
\begin{equation} \label{Ft}
F(t) = \left\{\begin{array}{cl} -F_0 \sin(\omega t)^2 \sin(\omega t)
    &\mbox{if }0<t<\frac{\pi}{\omega}\\0&\mbox{otherwise,}\end{array}\right.
\end{equation}
which has an amplitude of $F_0$, frequency $\omega = 0.05811$, and starts at
time $t=0$. Compared with \cite{BackProp2}, this field belongs to a half-cycle
pulse, $N=1/2$. The corresponding intensities are considered in the observed
regime. We will use the form (\ref{Ft}) in our examples, and later on comment
on some of the differences compared with (\ref{Ecos}).

We use this model in order to probe different definitions of the time when the
electron exits the tunnel.  The standard definition of tunneling exit points
equates the energy of the particle with the potential, at which time a
classical turning point would be reached in the absence of quantum
corrections. As shown in \cite{BackProp,BackProp2,BackProp3}, this condition
cannot always be imposed in non-static situations, in which the energy of the
electron is not constant and may not be known in an experiment. As an
alternative, these papers proposed classical back-propagation as a new method,
combined with a definition of the tunneling exit as the time when the momentum
of the particle in the direction of the force, evaluated on a classically
back-propagated trajectory, is zero.  However, while this condition is of
advantage in evaluations of experimental results \cite{BackProp2,BackProp3},
it is questionable, as also pointed out in \cite{Bohmian}, because it makes
use of a classical property (zero longitudinal momentum at a classical turning
point) in a region where classical physics is known to be inadequate. Our
methods describe tunneling by a single quantum trajectory, which we will
compare directly with the back-propagated classical trajectory in order to see
possible deviations.

\subsection{Definition of tunneling time for dynamic fields}

The main quantity of conceptual interest is called ``tunneling traversal
time'' in \cite{Bohmian}, which is the time the electron spends in a
classically forbidden region between two turning points. In a constant field,
the positions of turning points depend only on the initial energy of the
electron and can be easily determined, but the definition is more difficult to
implement when the dynamical behavior of the force is crucial
\cite{BackProp2}. 

\begin{figure}[htbp]
\begin{center}
\includegraphics[scale = 0.7]{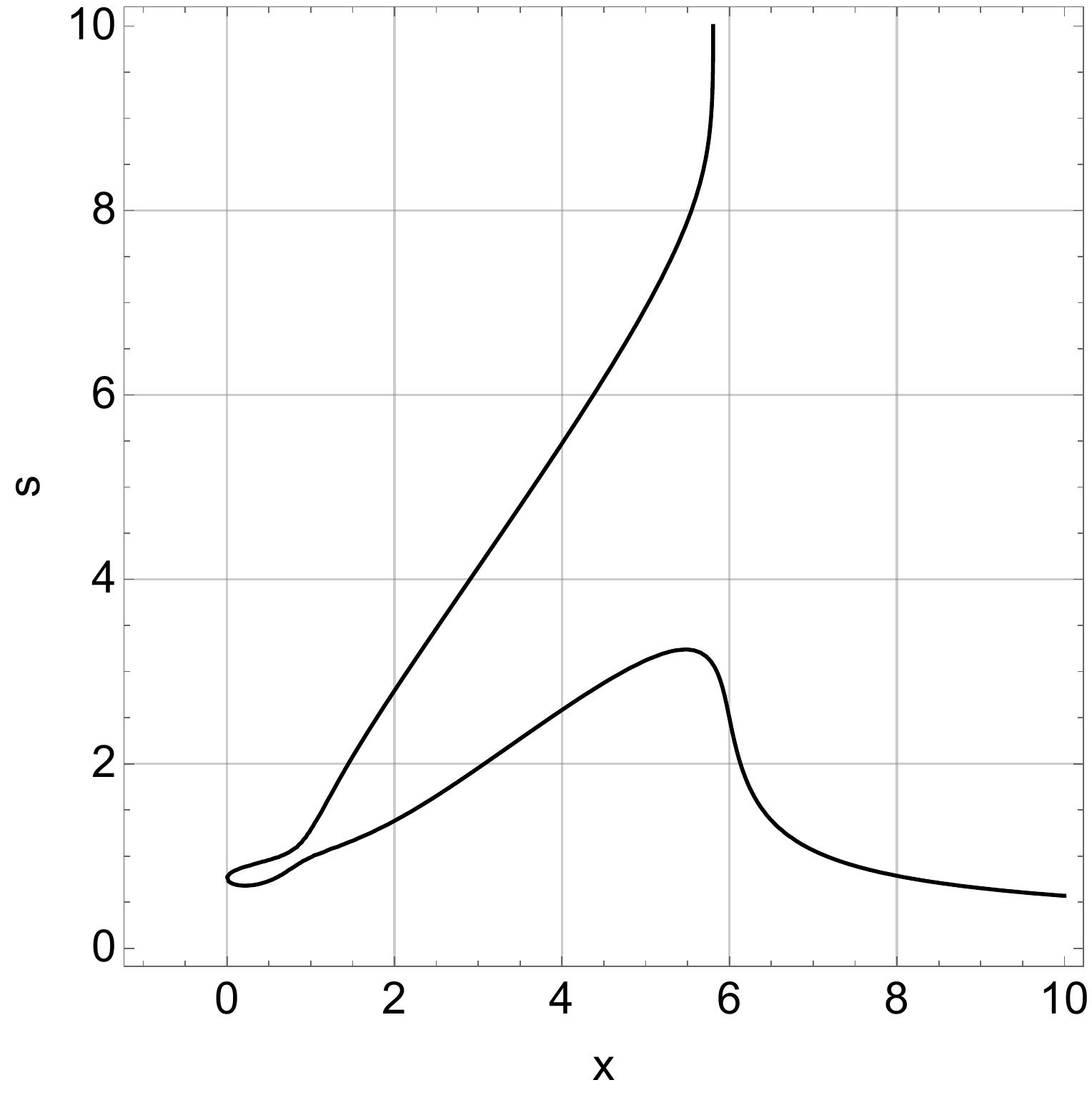}
\caption{Equipotential plot of the all orders at potental at $t=16$, about
  half-way to the wave peak. \label{fig:equipotential}}
\end{center}
\end{figure}

\begin{figure}
\begin{center}
 \includegraphics[scale = 0.8]{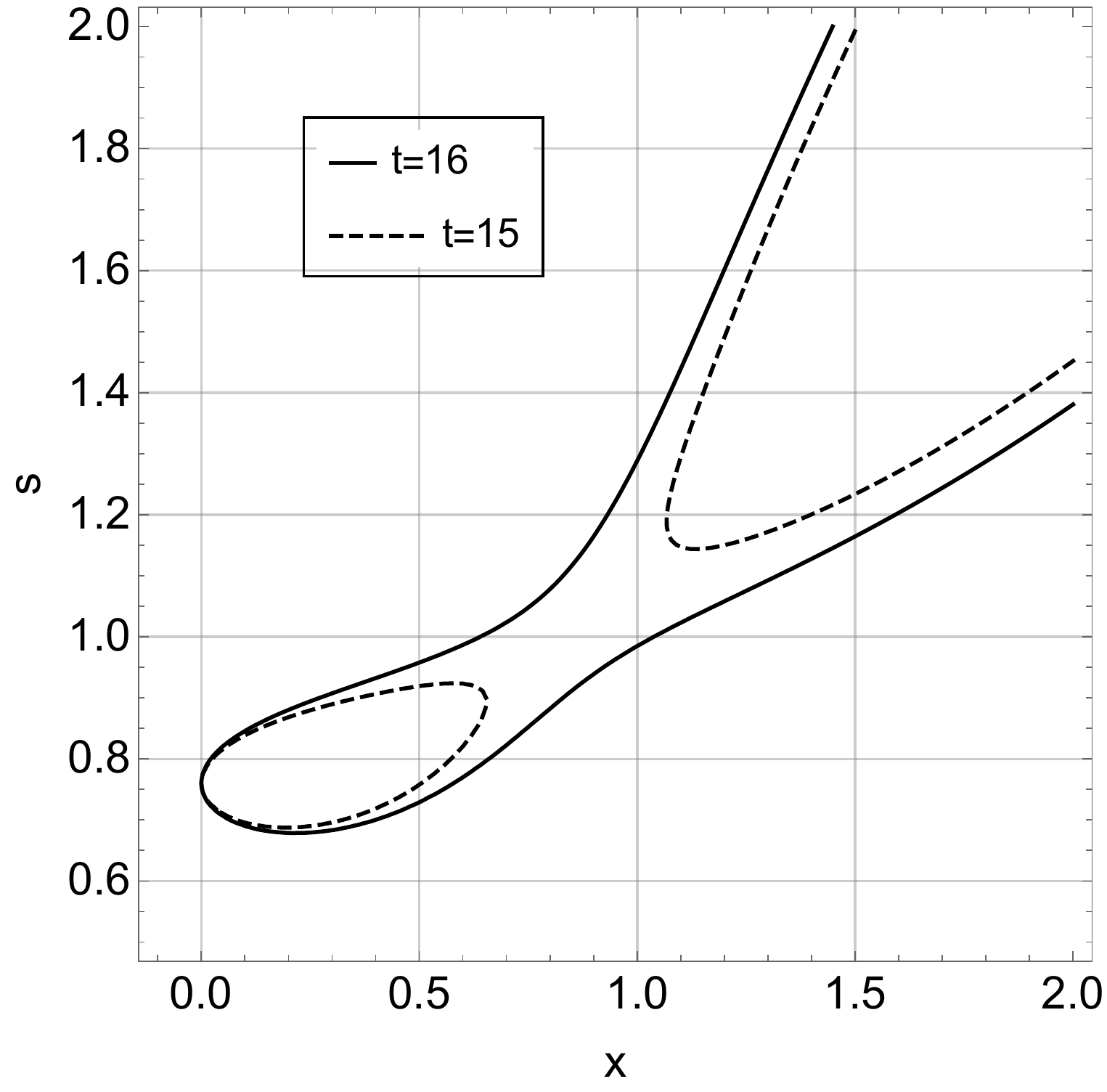}
 \caption{Zoom-in of Fig.~\ref{fig:equipotential} on the area of interest. The
   dashed contour is from $t=15$ at which time the tunneling channel has not
   completely opened. A little while later, at $t=16$, the tunneling channel
   is open and the particle can leave. \label{fig:ContourZoom}}
 \end{center}
\end{figure}

As a solution, \cite{BackProp2} proposed the method of classical
back-propagation in order to determine the ``tunneling exit time'' defined as
the point in time when the electron reenters a classically allowed region. By
definition, the tunneling exit time is therefore a point in time, while the
tunneling traversal time is a duration. The examples considered in
\cite{BackProp2} suggested near-zero tunneling exit times, which has to be
interpreted in the context of the pulse (\ref{Ecos}) with maximum
intensity at time zero. In the terminology of \cite{Bohmian}, the tunneling
exit time of \cite{BackProp2} is therefore equal to the ``tunneling ionization
time'' defined as the duration between the maximum of the external force and
the time when the electron reenters a classically allowed region.

The tunneling ionization time can be accessed in observations more directly
than the tunneling traversal time. But it does not give us a full picture of
the tunneling process because the electron may well start tunneling before the
external force has reached its maximum. The near-zero tunneling exit times or
tunneling ionization times of \cite{BackProp2} therefore do not imply that the
electron tunnels without any delay. The example of tunneling times given in
\cite{Bohmian} illustrates this difference, which we can show explicitly
using our effective potential: As shown in Figs.~\ref{fig:equipotential} and
\ref{fig:ContourZoom}, the tunnel has already opened as early as halfway
through the build-up of the external force.

In the next subsection, we will analyze tunneling exit criteria, and then
return to the question of tunneling traversal.

\subsection{Tunneling exit criteria for dynamic fields}

For the time-dependent potential \eqref{eq:Gaussian} we should use a
definition of tunneling exit time which can account for non-adiabatic
effects. For instance, the energy condition
\begin{equation}
\label{Tunnelingcriterion}
H_{\rm Q}(p(t),p_s(t),x(t),s(t);t) - x(t) \, F(t) = 0 
\end{equation}
gives us a finite time because we always have $V_{\rm eff}<0$ when the term
$U/2ms^2$ can be ignored. This definition focuses on the energy gain in an
external force: By the time the electron reaches zero energy, it is in an
allowed region for any negative potential. In this condition, quantum effects
can be significant, for instance when the kinetic energy $p_s^2/2m$ of
fluctuations raises the energy to positive values; see
Fig.~\ref{fig:PotentialEnergy} below. The condition is adapted to
non-adiabatic situations, in the sense that the dynamically changing energy is
kept track of.  While this criterion includes non-adiabatic effects, the
quantum dynamics is approximated by an all orders Hamiltonian.  The canonical
tunneling exit time is taken to be the instant when \eqref{Tunnelingcriterion}
is satisfied.

\begin{figure}[htbp]
\begin{center}
\includegraphics[scale=0.7]{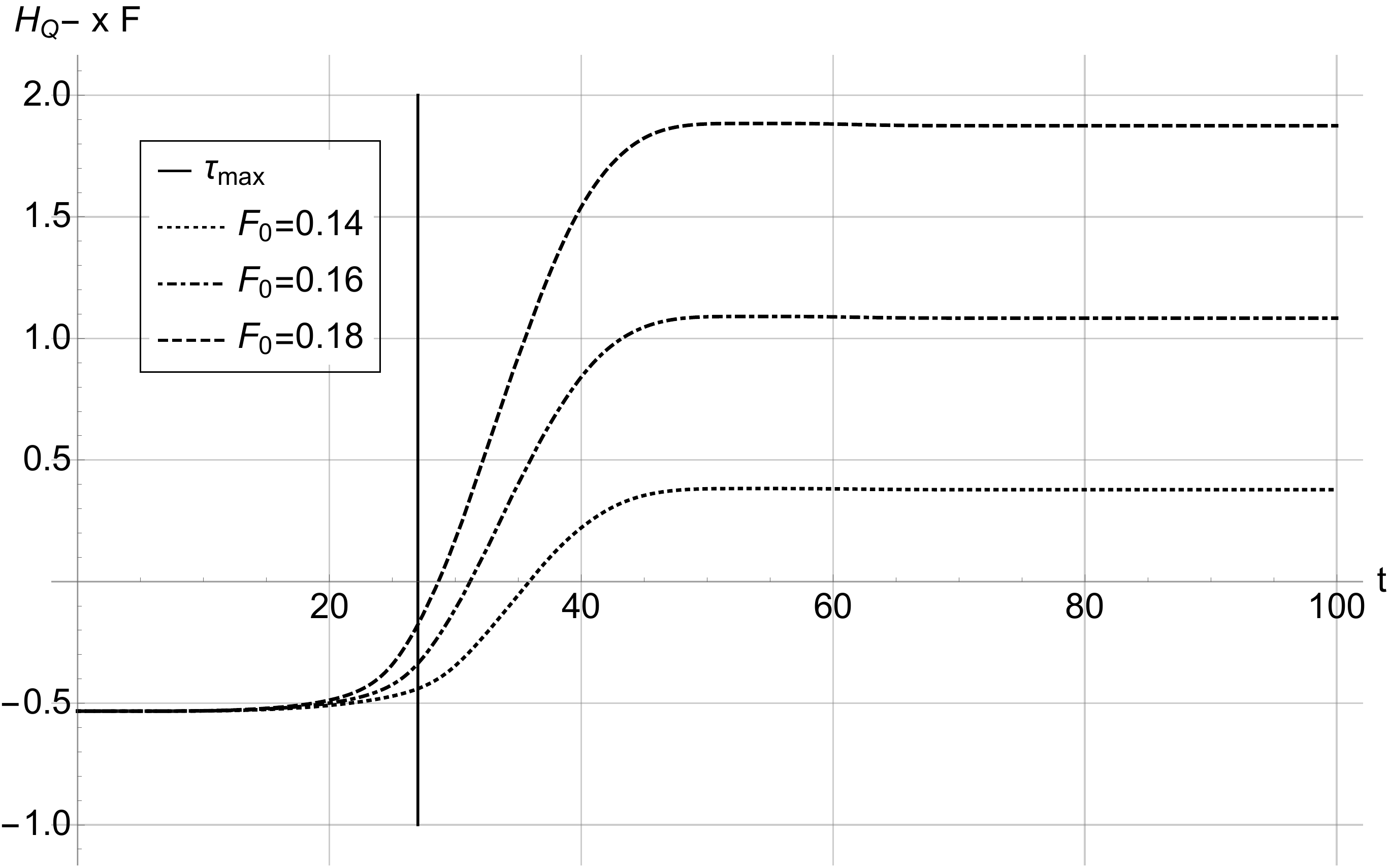}
\caption{The tunneling exit time as an energy condition: $H_{\rm Q} - x \,
  F=0$. The intermittent lines represent this condition with respect to time
  parameter $t$ for three different electric field amplitudes (corresponding
  to an intensity range of $F_0^2 \sim [6 \times 10^{14},12 \times 10^{14}] \,
  {\rm W/cm^2}$). The vertical solid line indicates the instant of maximum
  field strength at $\tau_{\mathrm{max}} \sim 27$ atomic units.}
\label{fig:energy}
\end{center}
\end{figure}

We present results from numerical simulations with the quantum Hamiltonian
(\ref{H}) for an effective potential (\ref{VEff3D}) and the initial conditions
\eqref{InitialValues}. We mainly show the tunneling exit time $\tau_{\rm ex}$
by extracting the instant when the interaction-free part of the quantum
Hamiltonian \eqref{Tunnelingcriterion} crosses the time axis. From this value,
we are able to determine the tunneling ionization time $\tau_{\rm ion} =
\tau_{\rm ex} - \tau_{\rm max}$, which is defined with respect to the instant
of maximum field, $t=\pi/2\omega$ in (\ref{Ft}); see Fig.~\ref{fig:energy}. In
particular, the tunneling ionization time $\tau_{\rm ion}$ is several atomic
units for a field amplitude $F_0=0.14$ and becomes smaller for higher
intensity pulses. Figure~\ref{fig:PotentialEnergy} shows that the ``quantum''
kinetic energy $T_Q = p_s^2/2m$ is important for an evaluation of this
condition. The tunneling exit time of the electron in Fig.~\ref{fig:energy}
explicitly indicates non-zero tunneling ionization time for a dynamic barrier,
similarly to what has been obtained in \cite{TunnelExperiment,TunnelWeakMeas,
  TunnelingResolution} but on a smaller scale.

\begin{figure}[htbp]
\begin{center}
\includegraphics[scale=0.7]{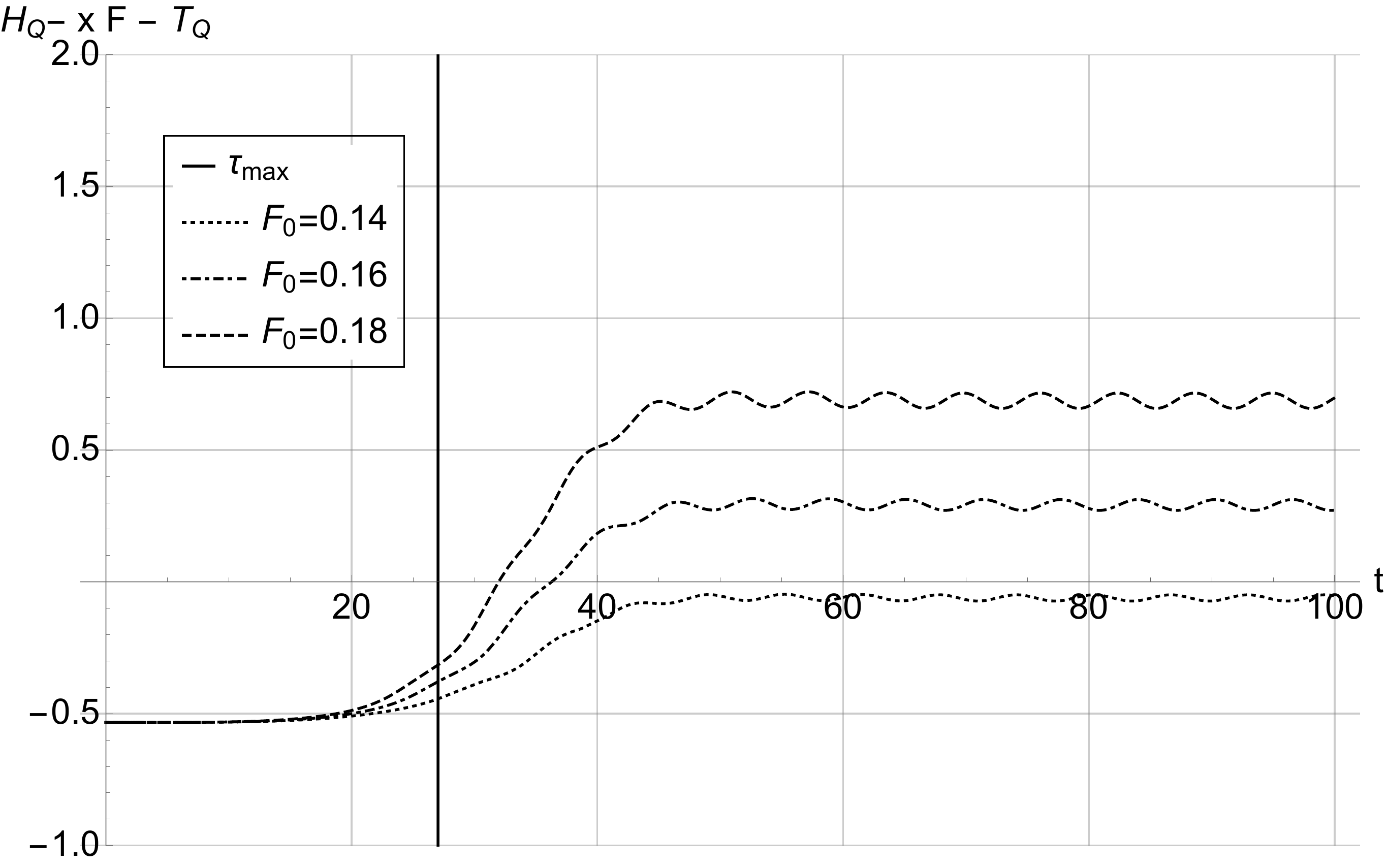}
\caption{The energy as a function of time, with the kinetic term of the
  quantum degrees of freedom removed.} 
\label{fig:PotentialEnergy}
\end{center}
\end{figure}

\begin{figure}[htbp]
\begin{center}
\includegraphics[scale=0.7]{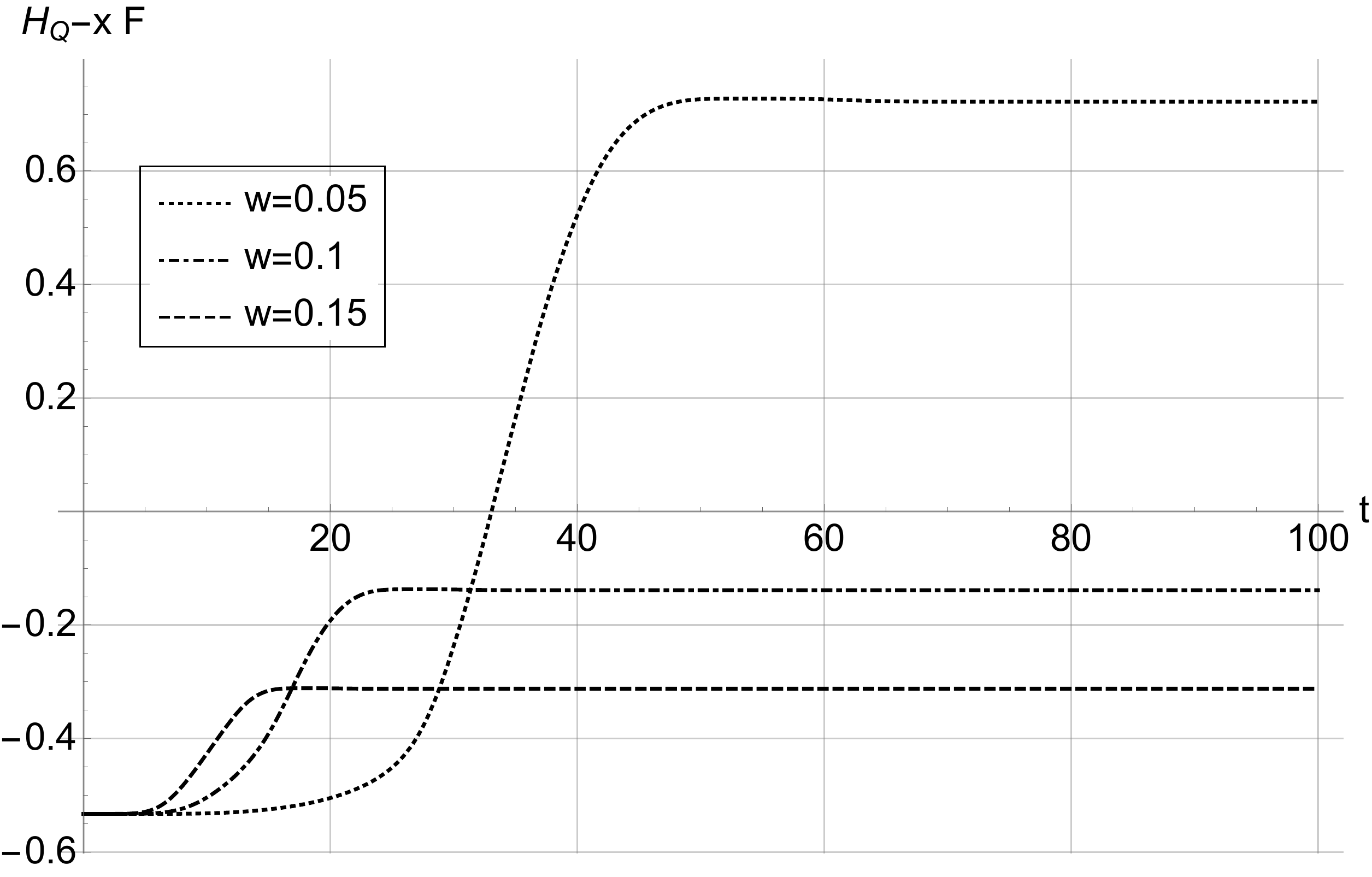}
\caption{Above a certain critical frequency we no longer obtain tunneling
  according to the condition (\ref{Tunnelingcriterion}).}
\label{fig:freqvary}
\end{center}
\end{figure}

In addition, laser pulses of sufficiently high frequency do not lead to
tunneling if we keep the same maximal field amplitude for varying frequencies;
see Fig.~\ref{fig:freqvary}. This implication is easy to understand because
less energy then falls on the atom. However, if we use pulses with various
frequencies and intensities such that there is always the same energy hitting
the atom, we find that, as the frequency rises, the tunneling exit criterion
gives ionization times that tend to zero.  In this limit, most of the energy
reaches the atom close to the wave peak. The result is conceptually similar to
the traditional distinction between tunneling ionization and multiphoton
ionization based on the Keldysh parameter $\gamma_{\rm K}=\omega\tau_{\rm K}$
with $\tau_{\rm K}=\sqrt{2|I_p|}/F$ \cite{Keldysh,Popov}. If $\gamma_{\rm
  K}\gg1$, the pulse frequency $\omega$ is too large to allow a process of
duration $\tau_{\rm K}$ to be completed during a laser cycle, which suggests
that tunneling does not take place at high frequency. However, the Keldysh
time $\tau_{\rm K}$ refers to the ionization potential $I_p$ and is therefore
adapted to a static electric field during tunneling.

Our method of approximating quantum dynamics allows us to compare different
possible tunneling criteria, in particular criteria based on momentum and
energy conditions for the tunnel exit. The recent study \cite{BackProp2},
analyzing a model for a single active electron in a helium atom, obtains a
near-zero ionization time using classical backpropagation and zero
longitudinal momentum to define the tunneling exit time. The basic idea of
classical backpropagation is to evolve the initial state quantum-mechanically
forward to some time after the laser pulse has ended. Then, the classically
transmitted ionized part of the wave packet is backpropagated and tunneling
exit properties are extracted corresponding to the specific tunneling
criterion applied.

 \begin{figure}[htbp]
           \begin{center}
 \includegraphics[scale = 0.7]{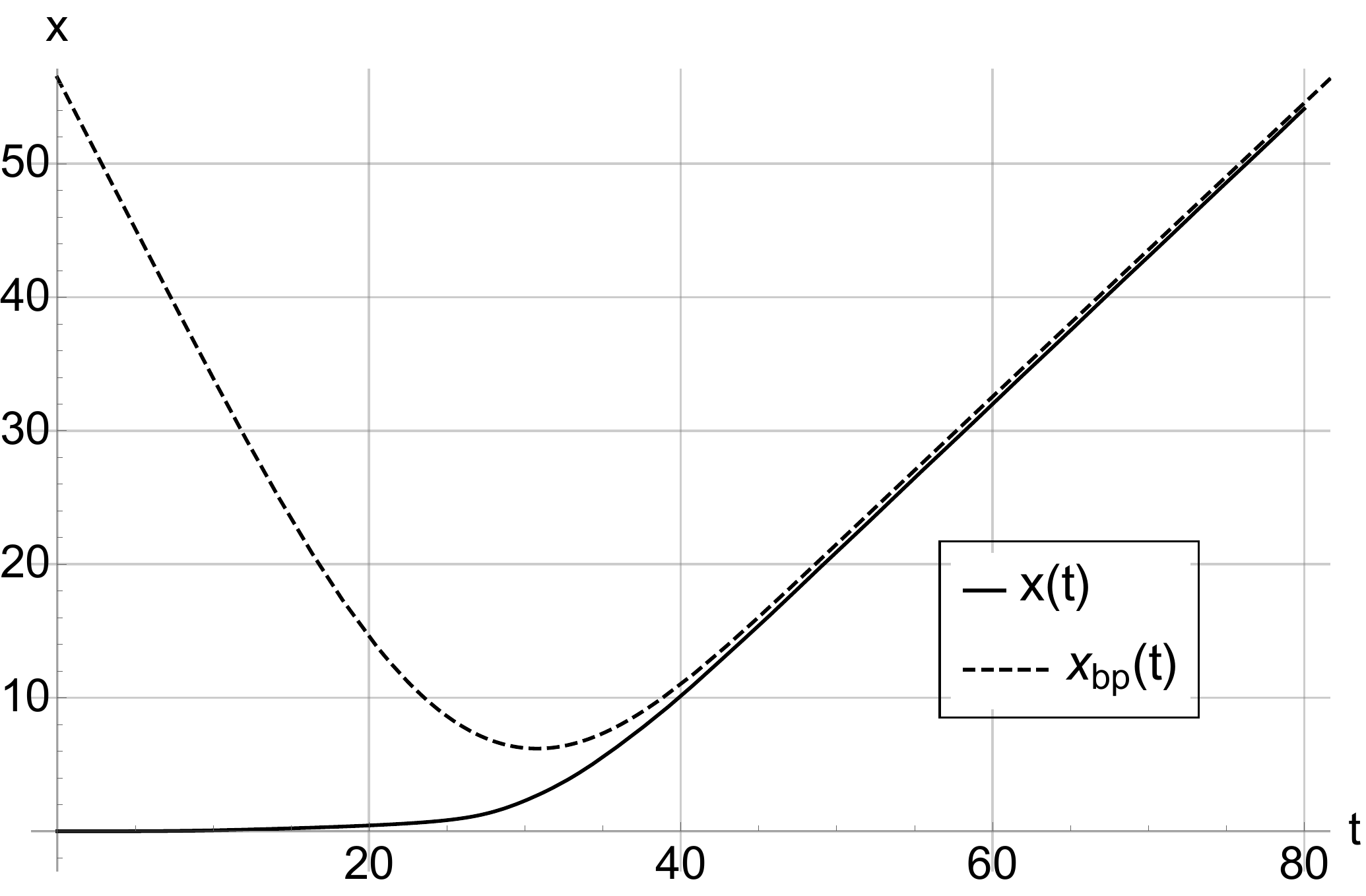}
\caption{Quantum trajectory (solid line) going forwards and the classical
  trajectory (dashed line) being back propagated in time. The quantum
  Hamiltonian is responsible for the evolution of the quantum trajectory. The
  back propagated trajectory is obtained by first evolving the classical
  trajectory backward in time with the initial condition of the quantum
  trajectory at some later time. \label{fig:backpropagation-trajectory}}
\end{center}
\end{figure}

\begin{figure}
\begin{center}
  \includegraphics[scale = 0.5]{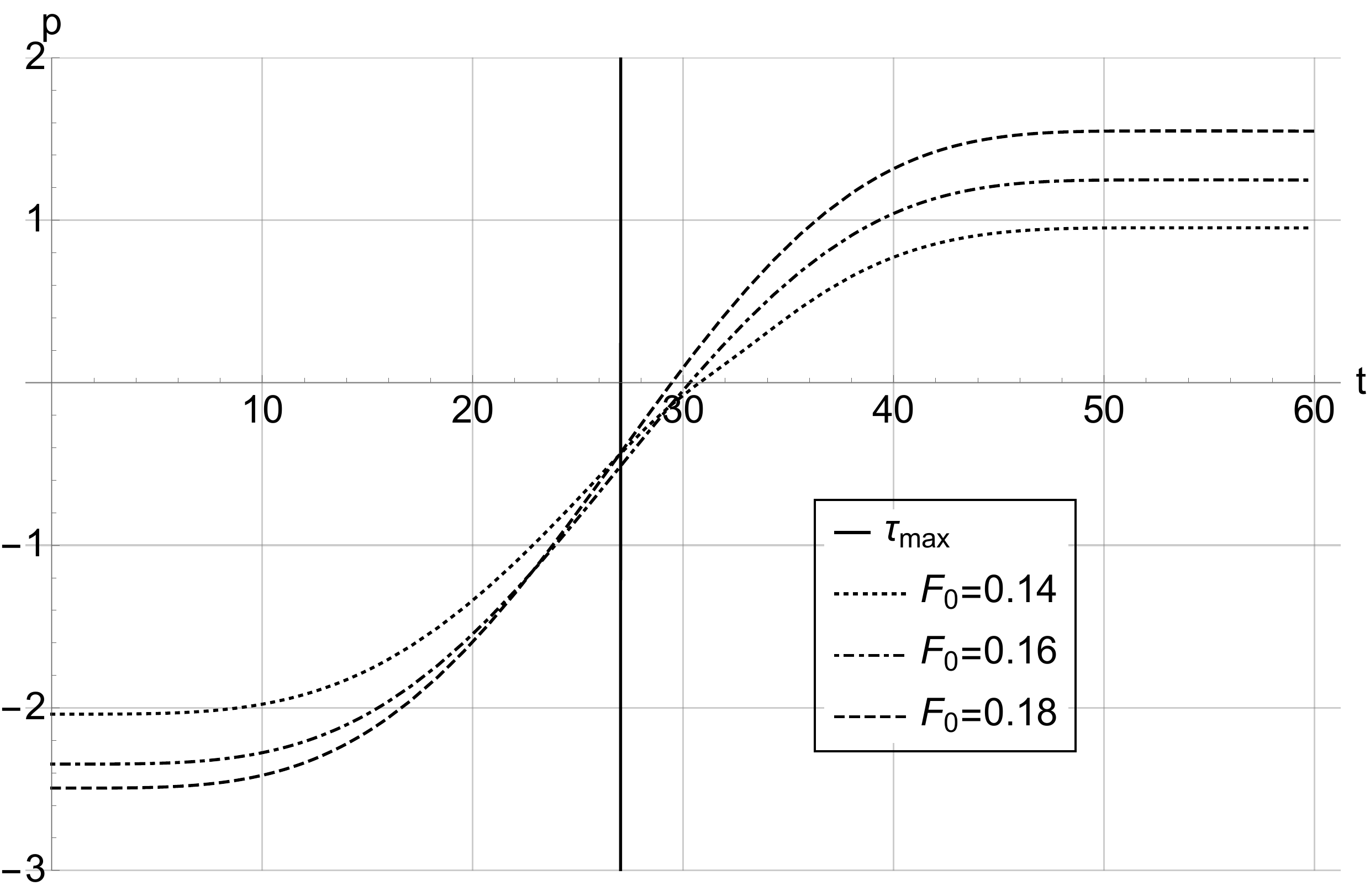}
  \caption{Momentum, as a function of time, being back propagated in time. The
  intermittent lines represent the momentum condition with respect to time
  parameter $t$ for the same three different electric field amplitudes used for
  the energy condition. The vertical line indicates the instant of maximum
  field strength $\tau_{\mathrm{max}} \sim 27$ atomic units.
  \label{fig:backpropagation-momentum}} 
    \end{center}
        \end{figure}

We can compare the momentum condition with the energy condition that we
introduced in \eqref{Tunnelingcriterion}. First, we evolve the system by the
quantum Hamiltonian in \eqref{H} forward to some late time, $t \sim
150$. Then, using the final values of $\{\langle \hat{x} \rangle, \langle
\hat{p} \rangle\}$ at the late time as initial conditions of position and
momentum $\{x_{\rm bp}, p_{\rm bp}\}$, we use the classical Hamiltonian
$H_{\rm cl} \equiv p^2/2 + V(x)$ to backpropagate classically to an early
time. Figure~\ref{fig:backpropagation-trajectory} shows that the
backpropagation trajectory of the particle stays rather close to the quantum
evolved trajectory. However, the backpropagated trajectory deviates from the
effective trajectory around the instant ($t\approx 27$) when the electric
field amplitude is maximum, close to the tunneling exit, where it bounces off
the potential well. In Fig.~\ref{fig:backpropagation-momentum} we show how the
tunneling exit time is realized with respect to the momentum condition based
on classical backpropagation. There is a non-zero tunneling ionization time
$\tau_{\rm ion} \sim 3$ (atomic units) in qualitative agreement with but
smaller than what we obtained from the energy condition.

\begin{figure}[htbp]
\begin{center}
\includegraphics[scale=0.55]{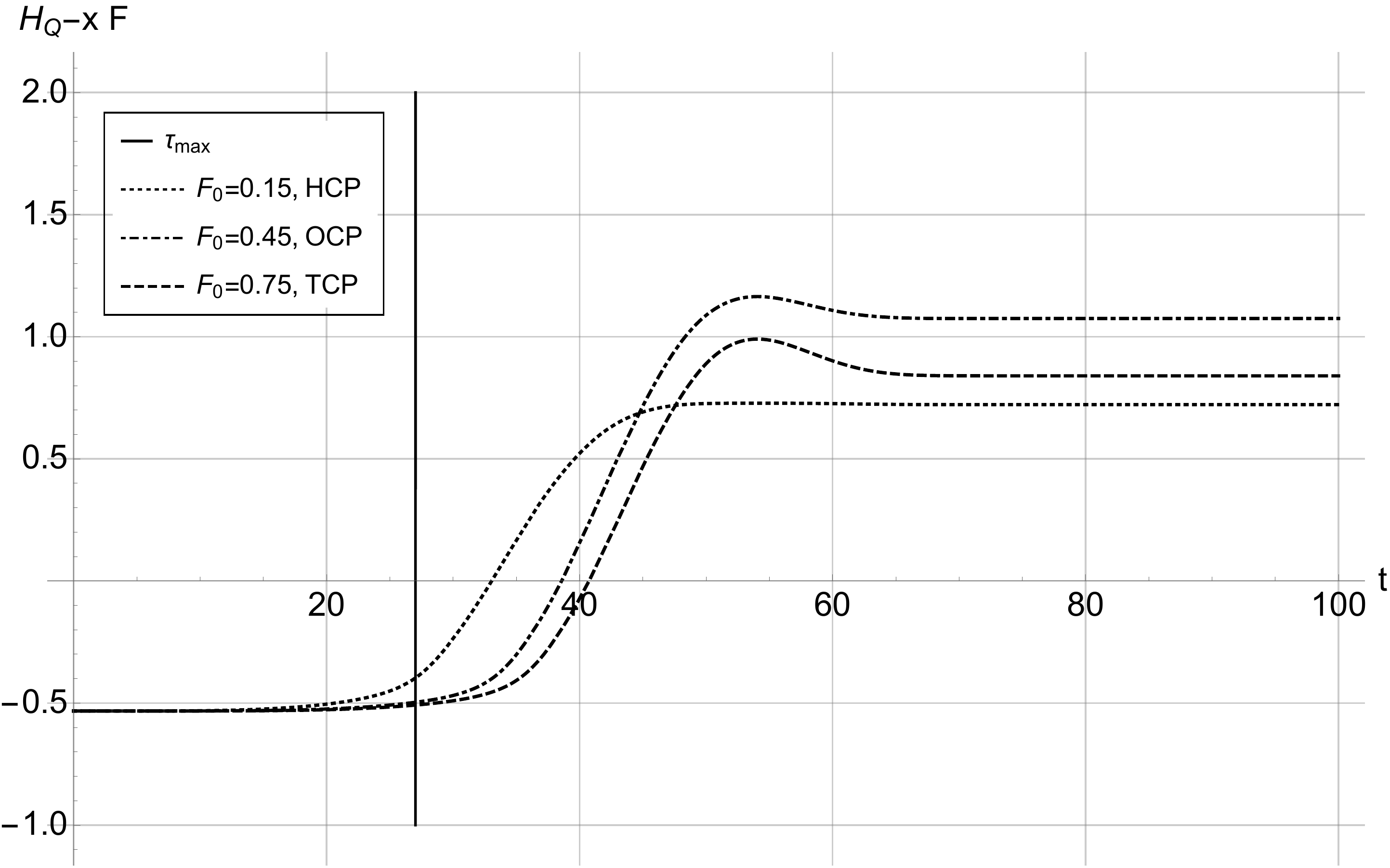}
\caption{The tunneling energy condition as a function of time for various
  pulses with different field amplitudes: HCP (half-cylce pulse,
  $F_0=0.15$), OCP (one-cycle pulse, $F_0=0.45$), TCP (two-cycle pulse,
  $F_0=0.75$).}
\label{fig:half-one-two}
\end{center}
\end{figure}

 \begin{figure}[htbp]
           \begin{center}
\includegraphics[scale = 0.7]{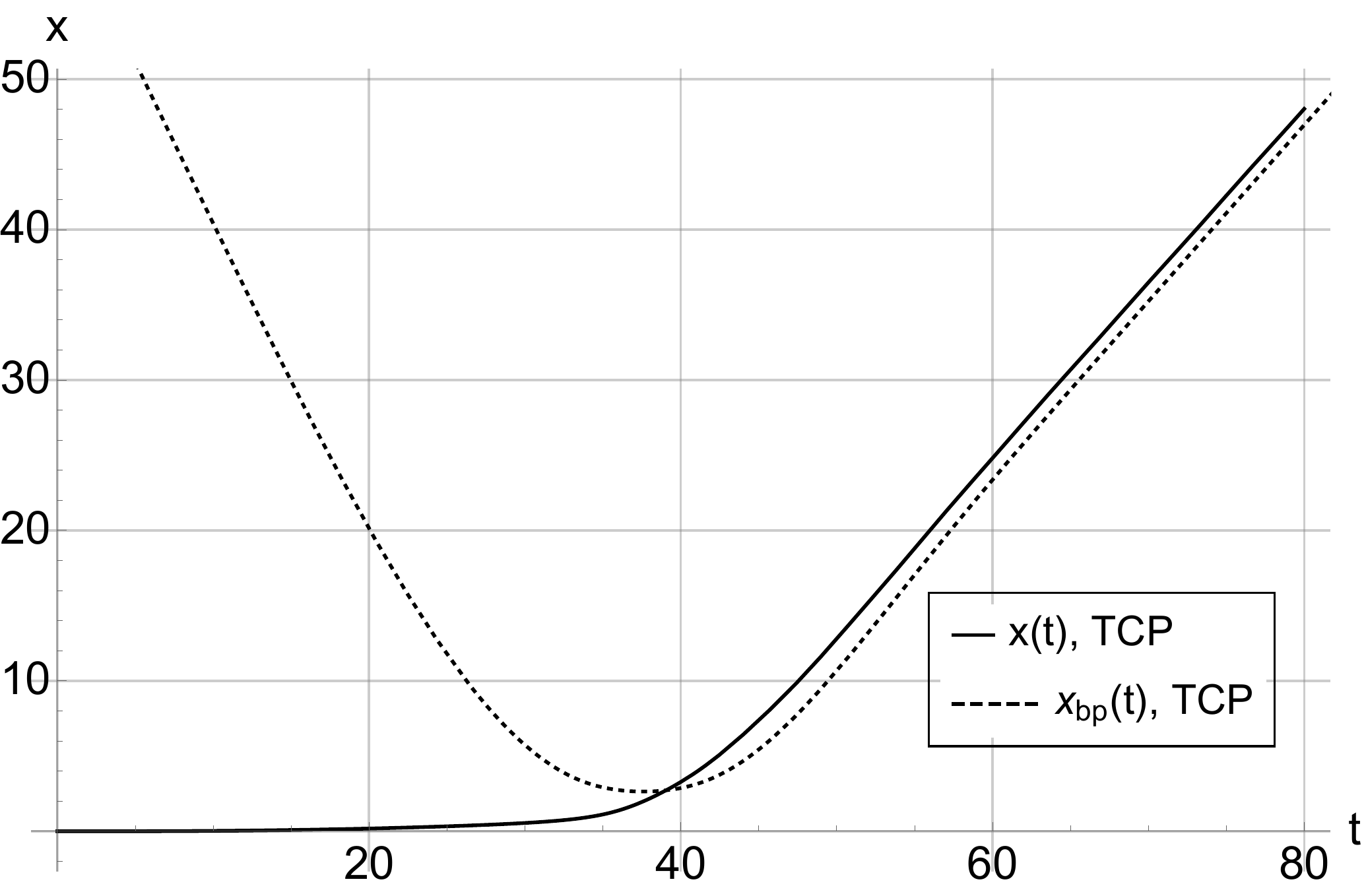}
\caption{Quantum trajectory (solid line) going forwards and the classical
  trajectory (dashed line) being back propagated in time for a two-cycle
  pulse. The quantum Hamiltonian is responsible for the evolution of the
  quantum trajectory. \label{fig:backpropagation-trajectory-two}}
\end{center}
\end{figure}

\begin{figure}
\begin{center}
\includegraphics[scale = 0.7]{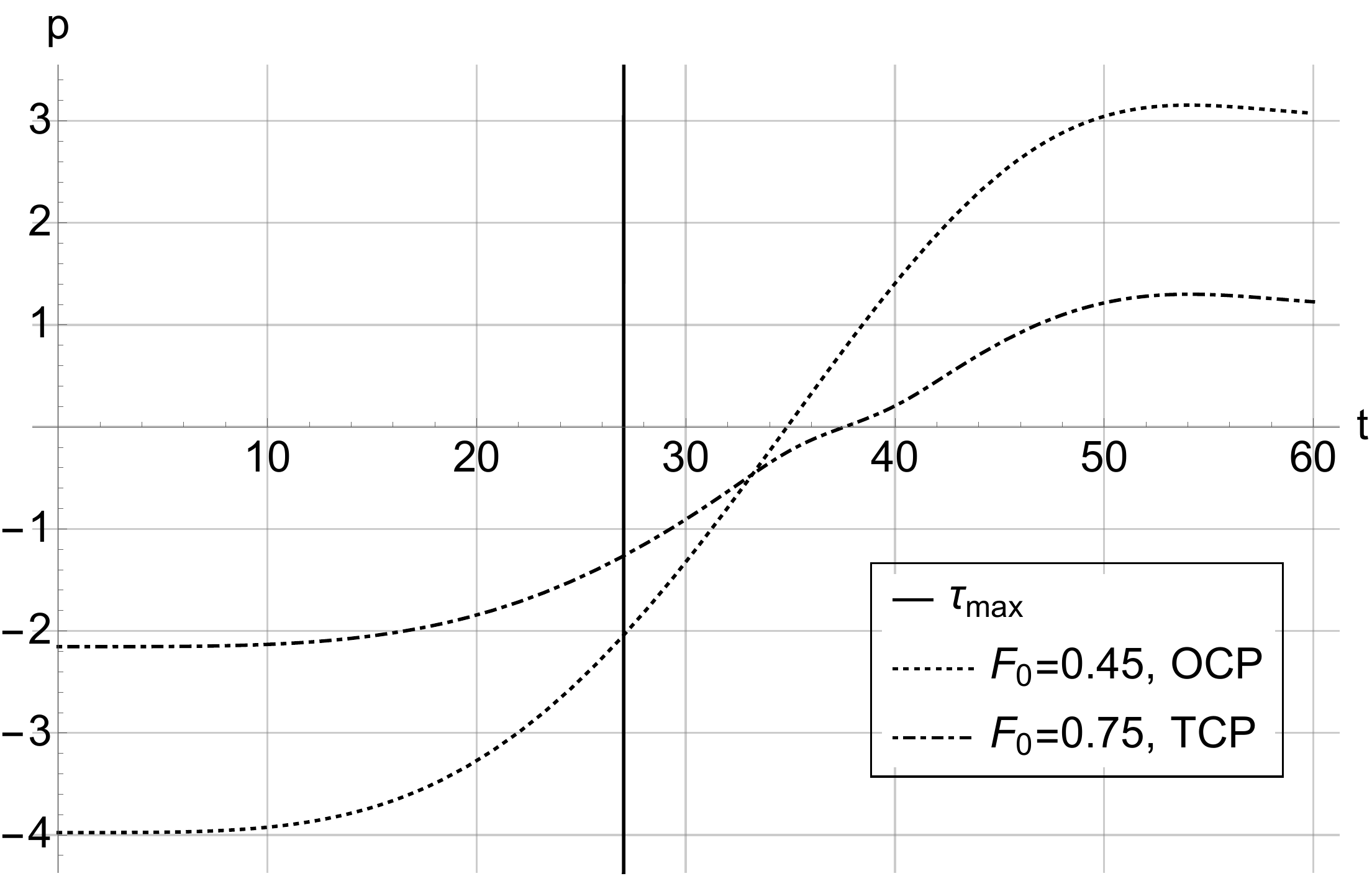}
\caption{Momentum, as a function of time, being back propagated in time for
  both one- and two-cycle pulses. The intermittent lines represent the
  momentum condition with respect to time parameter $t$ for the same three
  electric field amplitudes used for the energy
  condition. \label{fig:backpropagation-momentum-one-two}}
  \end{center}
\end{figure}
        
So far, our results have been shown for a half-cycle pulse (\ref{Ft}), while
\cite{BackProp2} used a two-cycle pulse. We repeated our calculations for one-
and two-cycle pulses while keeping the same frequency used in the half-cycle
pulse, see
Fig.~\ref{fig:half-one-two}. Figures~\ref{fig:backpropagation-trajectory-two}
and \ref{fig:backpropagation-momentum-one-two} confirm our general findings,
and they show that tunneling is possible for significantly larger field
amplitudes than for a half-cycle pulse (for which less energy falls on the
atom). The frequency dependence of tunneling times can also be confirmed. More
cycles in a pulse of the same frequency produce a longer tunneling ionization
time according to both criteria evaluated here because the field intensity
rises more slowly for bigger $N$.

\subsection{Tunneling dynamics of Hydrogen in three dimensions}

As the most realistic one of our models, we now consider the three dimensional
case of a Hydrogen atom in a time dependent electric field
\begin{equation}\label{ClassHam}
H=\frac{1}{2}|\vec{p}|^2-\frac{1}{|\vec{r}|}+\vec{r}\cdot \vec{E}(t) \, ,
\end{equation}
where we use a half cycle pulse
\begin{equation}
  \vec{E}(t)=-E_0 \sin^2{(\omega t)}\theta(t)\theta(\pi/\omega-t)
  \left(\begin{array}{c} \sin{(\omega 
        t)}\\\cos{(\omega  t)}\\0\end{array}\right)
  \, . 
\end{equation}
The classical Hamiltonian at (\ref{ClassHam}) has the all-orders quantization
given in (\ref{VEff3D}).

\begin{figure}[htbp]
\begin{center}
\includegraphics[scale=0.7]{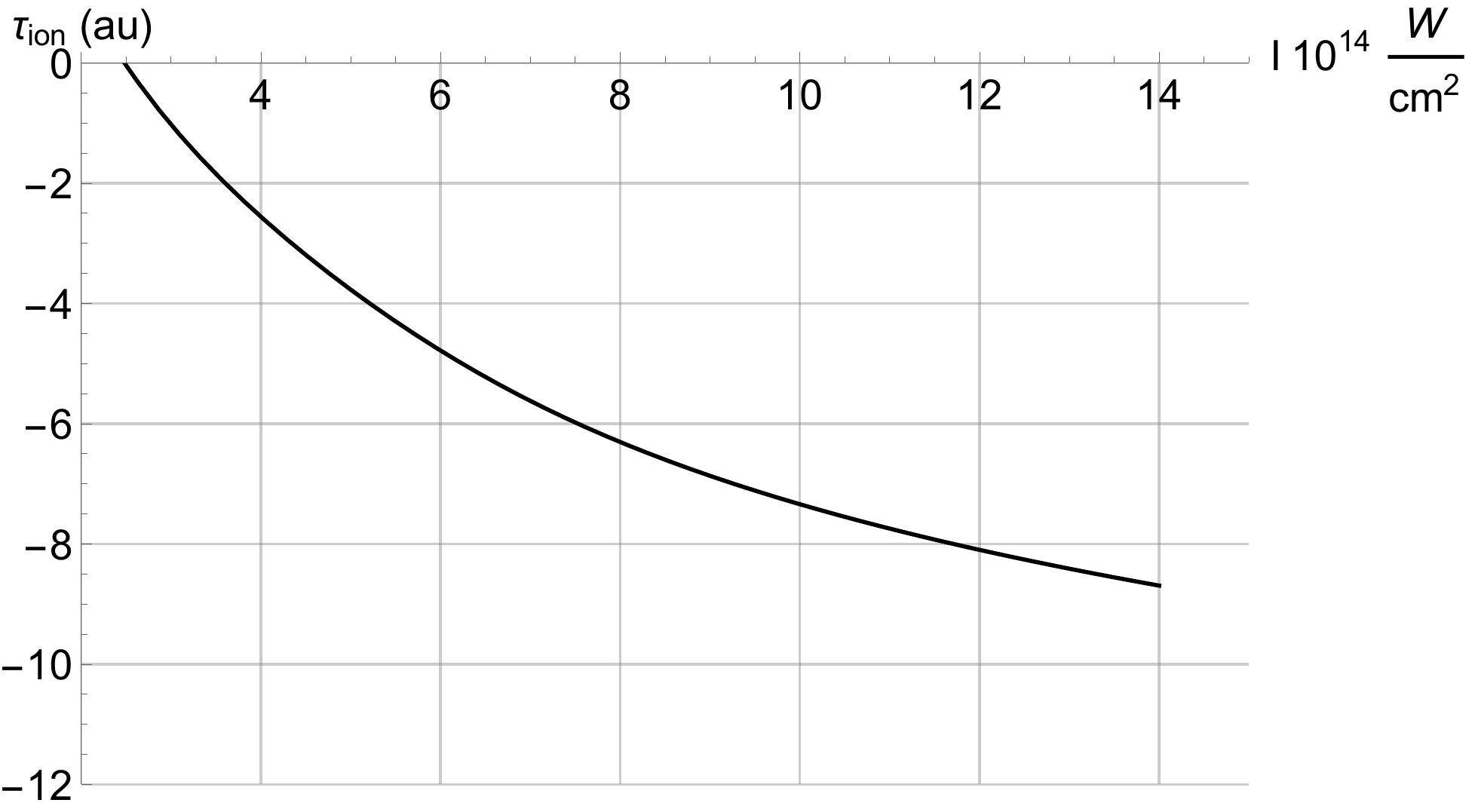}
\caption{Ionization time as a function of the laser intensity in the
  3-dimensional model (\ref{ClassHam}). } 
\label{fig:ion_time}
\end{center}
\end{figure}
        
We use the definition of the tunnel exit time as the moment when the quantum
Hamiltonian with the electric field term removed is zero: $H_Q-\vec{r}\cdot
\vec{E}(t)=0$. The ionization time is then defined as the difference between
the time of the maximum electric field strength and the exit time, $\tau_{\rm
  ion}=\tau_{\rm ex}-\tau_{\rm max}$, and shown in
Fig.~\ref{fig:ion_time}. Depending on the peak laser intensity, we find an
ionization time that is either positive or negative.  We can easily understand
this result as showing that the electron can tunnel well before the peak
reaches the atom, provided the intensity of the pulse is large
enough. However, a negative ionization time does not imply that there is no
tunneling delay.
 
\begin{figure}[htbp]
           \begin{center}
  \includegraphics[scale = 0.7]{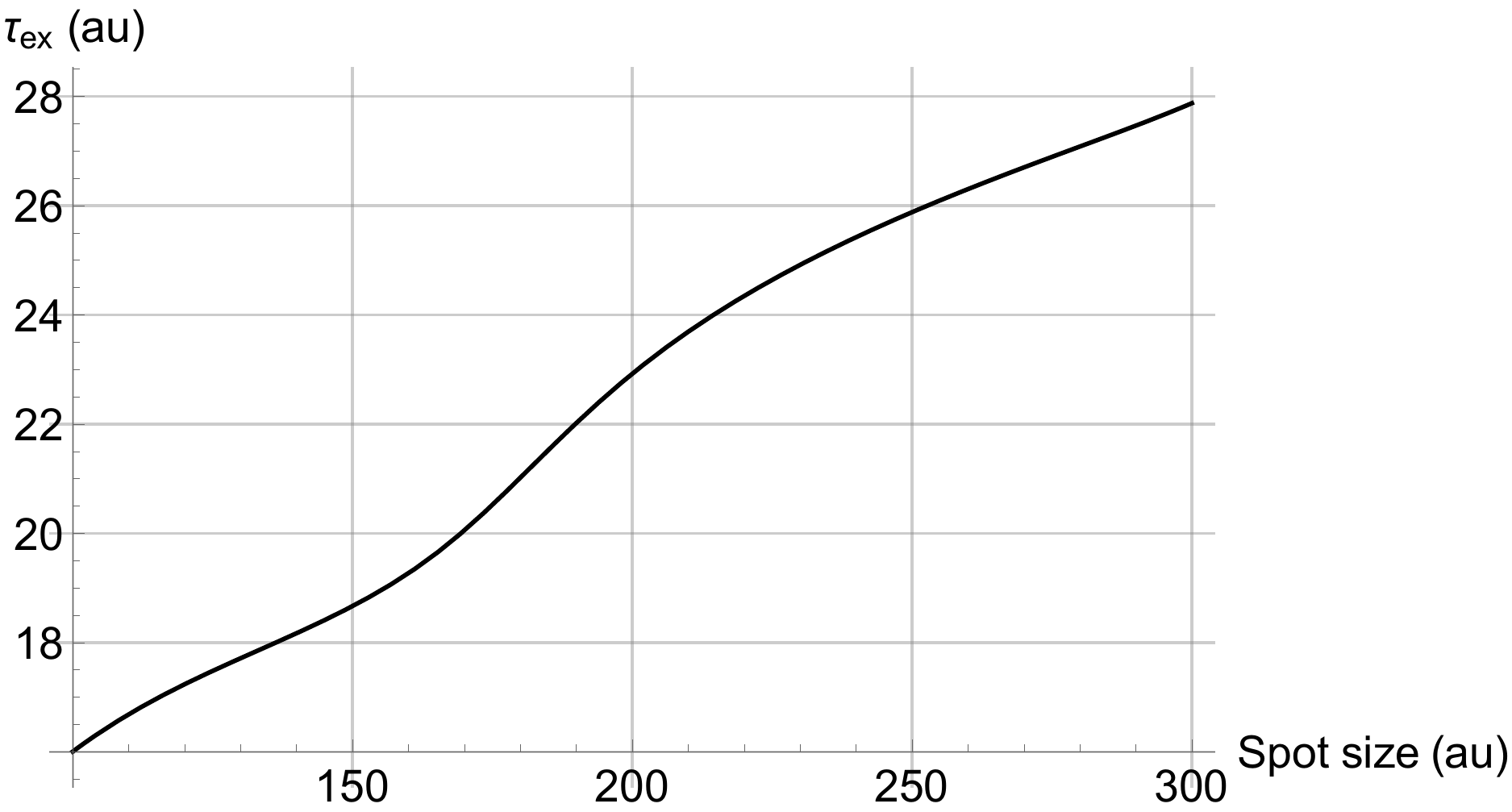}
\caption{The exit time as a function of the spot size at a distance of 1000
  atomic units. \label{fig:spotvtime}}
 \end{center}
\end{figure}

\begin{figure}
\begin{center}
\includegraphics[scale = 0.7]{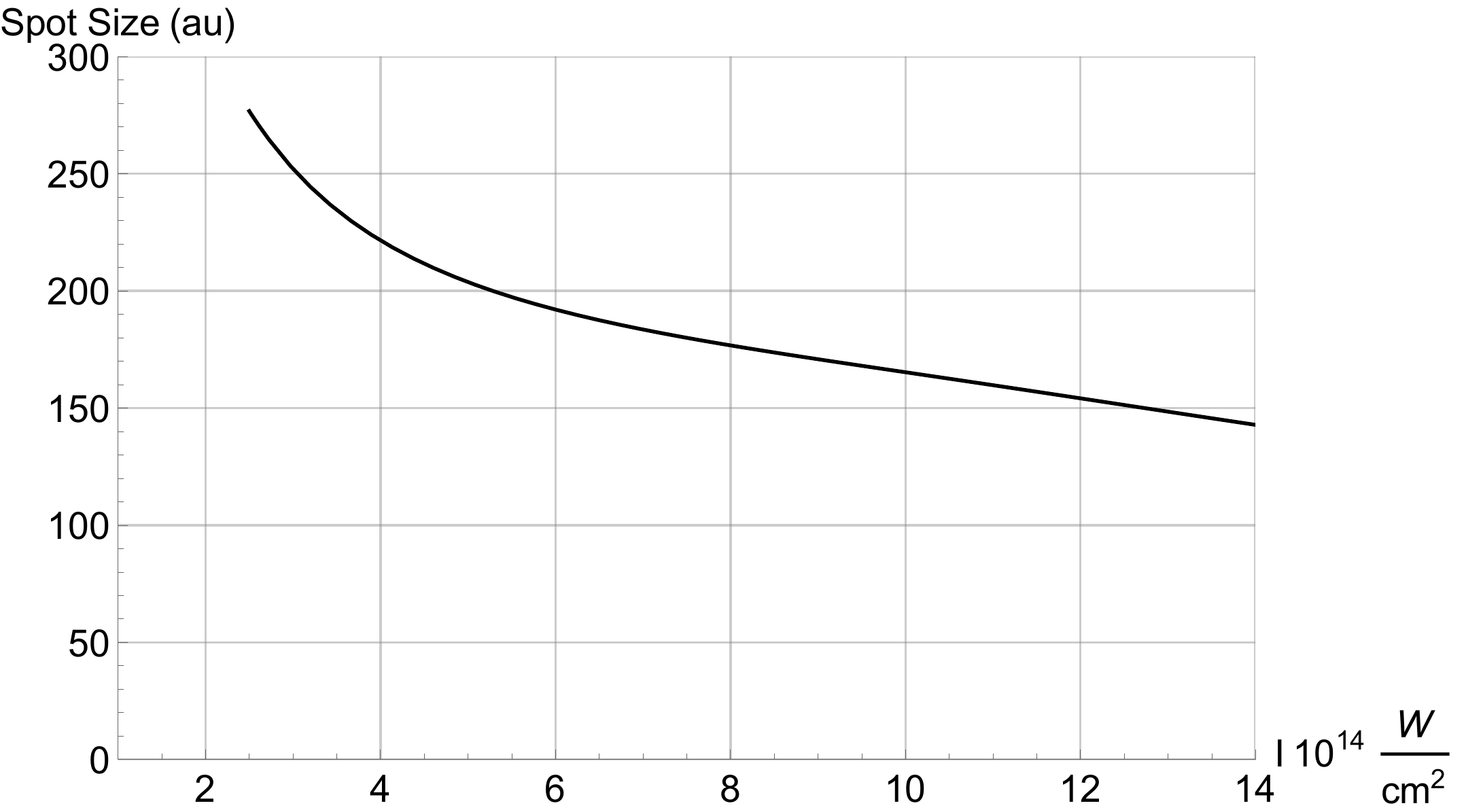}
              \caption{Spot size of the wave packet a distance of 1000 atomic
                units from the 
  atom. \label{fig:spotsizeintensity}}
                       \end{center}
\end{figure}
        
\begin{figure}[htbp]
           \begin{center}
\includegraphics[scale = 0.8]{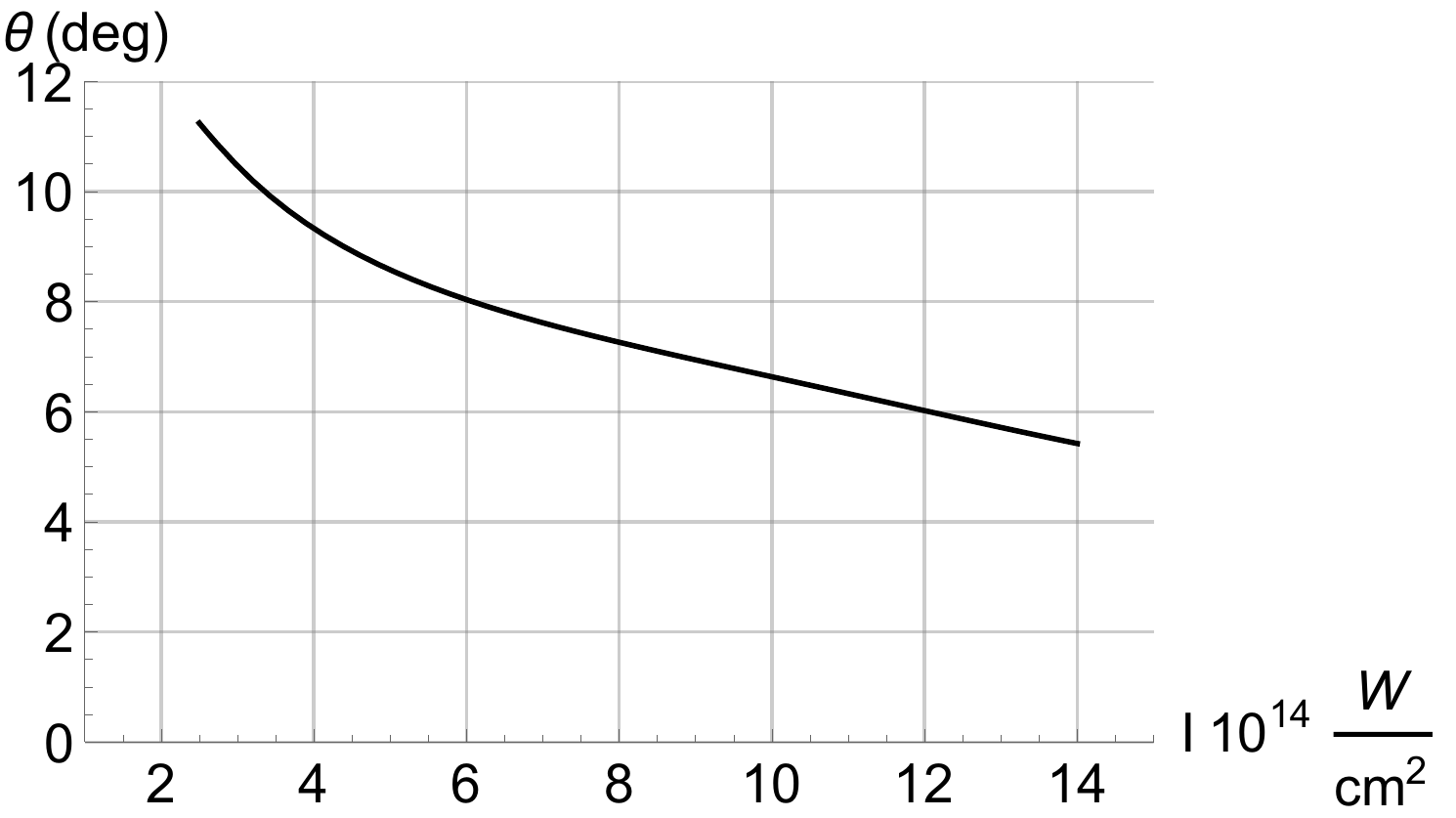}
\caption{Off-set angle of the ionized part of the wave
  packet. \label{fig:offtsetvsI}} 
\end{center}
\end{figure}

\begin{figure}
\begin{center}
\includegraphics[scale = 0.8]{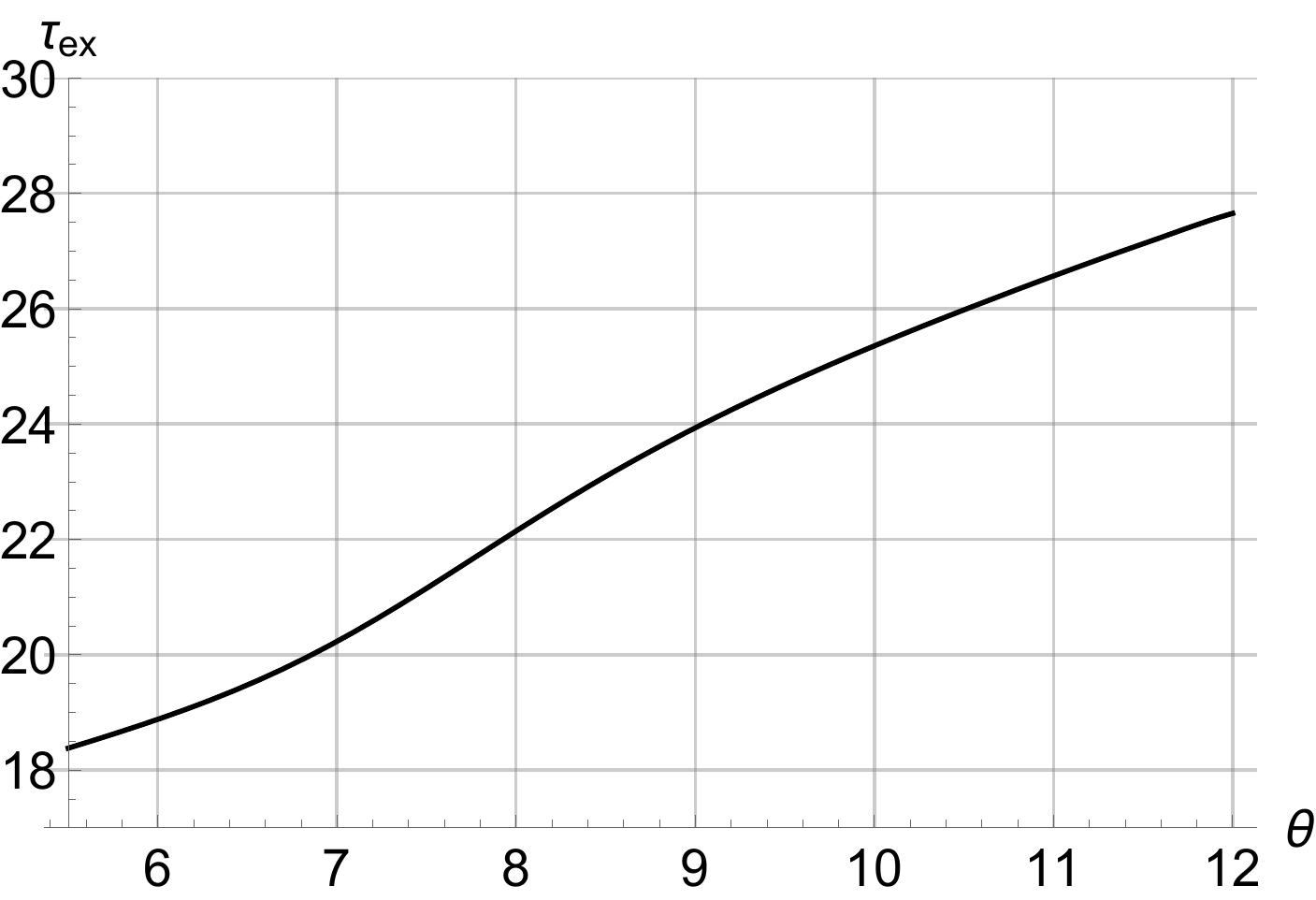}
  \caption{Tunneling exit time in terms of the offset
    angle. \label{fig:offsetvstau}} 
                       \end{center}
        \end{figure}

Other observables are also accessible as well as correlations between
them. Figures~\ref{fig:spotvtime} and \ref{fig:spotsizeintensity} show that
the spot size of the electron jet, defined as the geometric mean of the
transversal fluctuations, depends monotonically on the exit
time. This result indicates that there is indeed a tunneling delay, or at
least non-trivial tunneling dynamics, even if the ionization time is negative:
The larger the exit time, the more time there is for the wave packet to spread
out. Additionally, the tunneling time depends monotonically on the offset
angle, see Figures~\ref{fig:offtsetvsI} and \ref{fig:offsetvstau}.
  
\subsection{Alternate Definition of tunneling time}

The transverse fluctuations used to define the spot size have an interesting
dynamics which can be used to define the tunneling exit time in an inherently
quantum way, rather than using classical dynamics as in backpropagation.  As
indicated by Fig.~\ref{Evolution}, and confirmed below for the 3-dimensional
non-static model, the transversal fluctuations have three phases. Initially,
the particle is confined for some time and the fluctuations stay constant near
their ground-state values. During tuneling in the second phase, the state and
its fluctuations undergo a more complicated dynamics. After tunneling and when
the pulse has ended, during the third phase transversal fluctuations grow
linearly as is well-known for a free particle. These phases are clearly
demarcated in a plot of the fluctuations, which are readily accessible from
simulations in our effective potential. 

Nevertheless, extracting the transverse fluctuations is not entirely
trivial. To do so, we transform to the co-rotating frame in which some
fluctuation parameters $s_i$ are transverse to the external force at all
times. Under global rotations, the second-order position moments of a state,
defined in general as
\begin{equation}
 \Delta_{ij}=
 \langle(\hat{r}_i-\langle\hat{r}_i)(\hat{r}_j-\langle\hat{r}_j)\rangle\,,
\end{equation}
transform in the following way
\begin{equation}
\bar{\Delta}_{ij}=\mathcal{O}_{k i}\Delta_{k l} \mathcal{O}_{l j}\,,
\end{equation}
where $\mathcal{O}_{ij}$ is the rotation matrix that acts on position
coordinates.  This transformation results in the transverse fluctuation
\begin{equation}
s_{\rm T}=\sqrt{\cos^2{(\theta(t))}s_x^2+\sin^2{(\theta(t))}s_y^2}
\end{equation}
where $\theta$ is the offset angle as a function of time. 

\begin{figure}[htbp]
           \begin{center}
  \includegraphics[scale = 0.7]{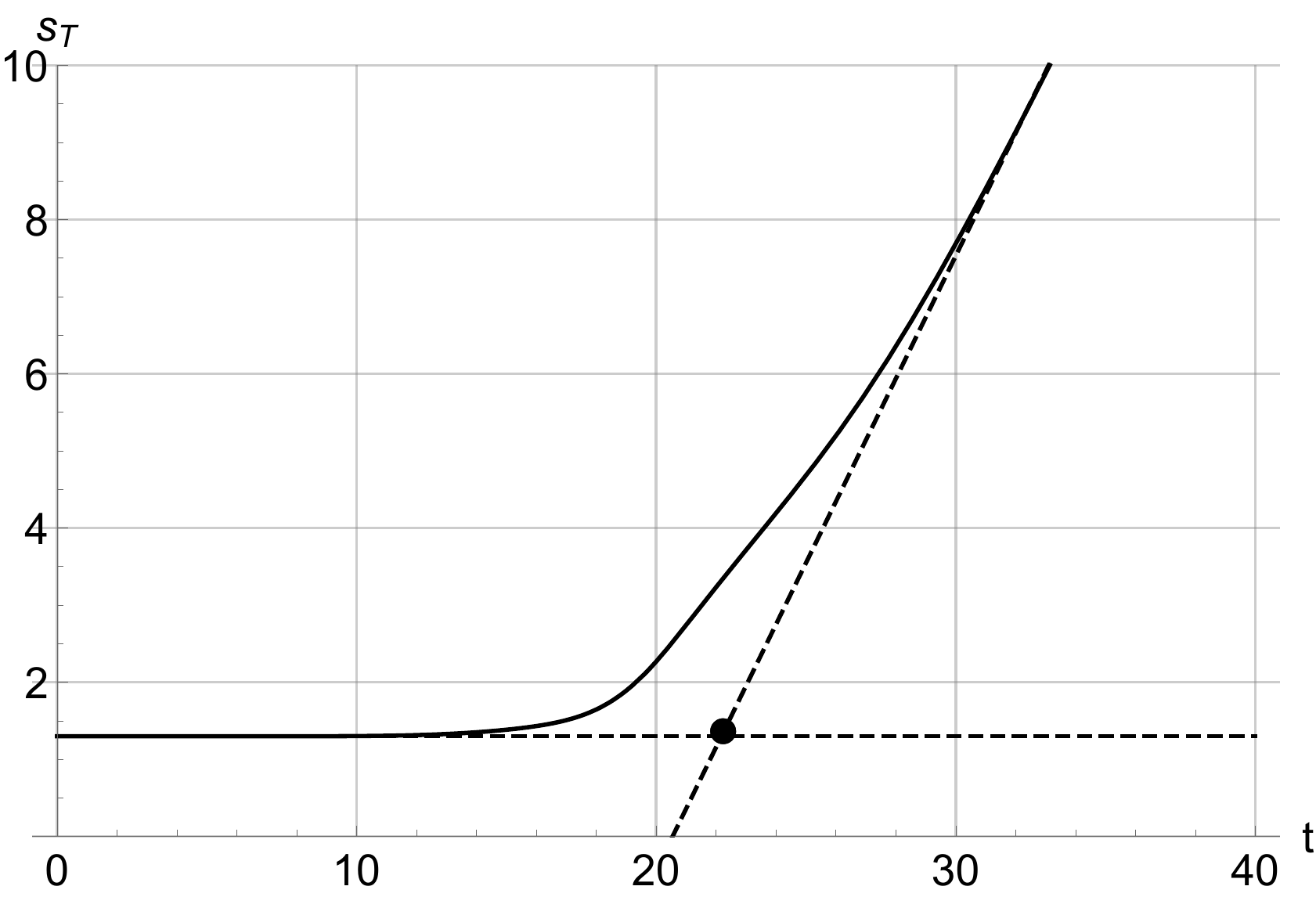}
\caption{The transverse fluctuations as a function of time. The
  tangent lines of the linear regions are plotted in the dotted lines, and
  their intersection is marked with a dot. \label{fig:trans_size}}
\end{center}
\end{figure}

\begin{figure}
\begin{center}
\includegraphics[scale=0.7]{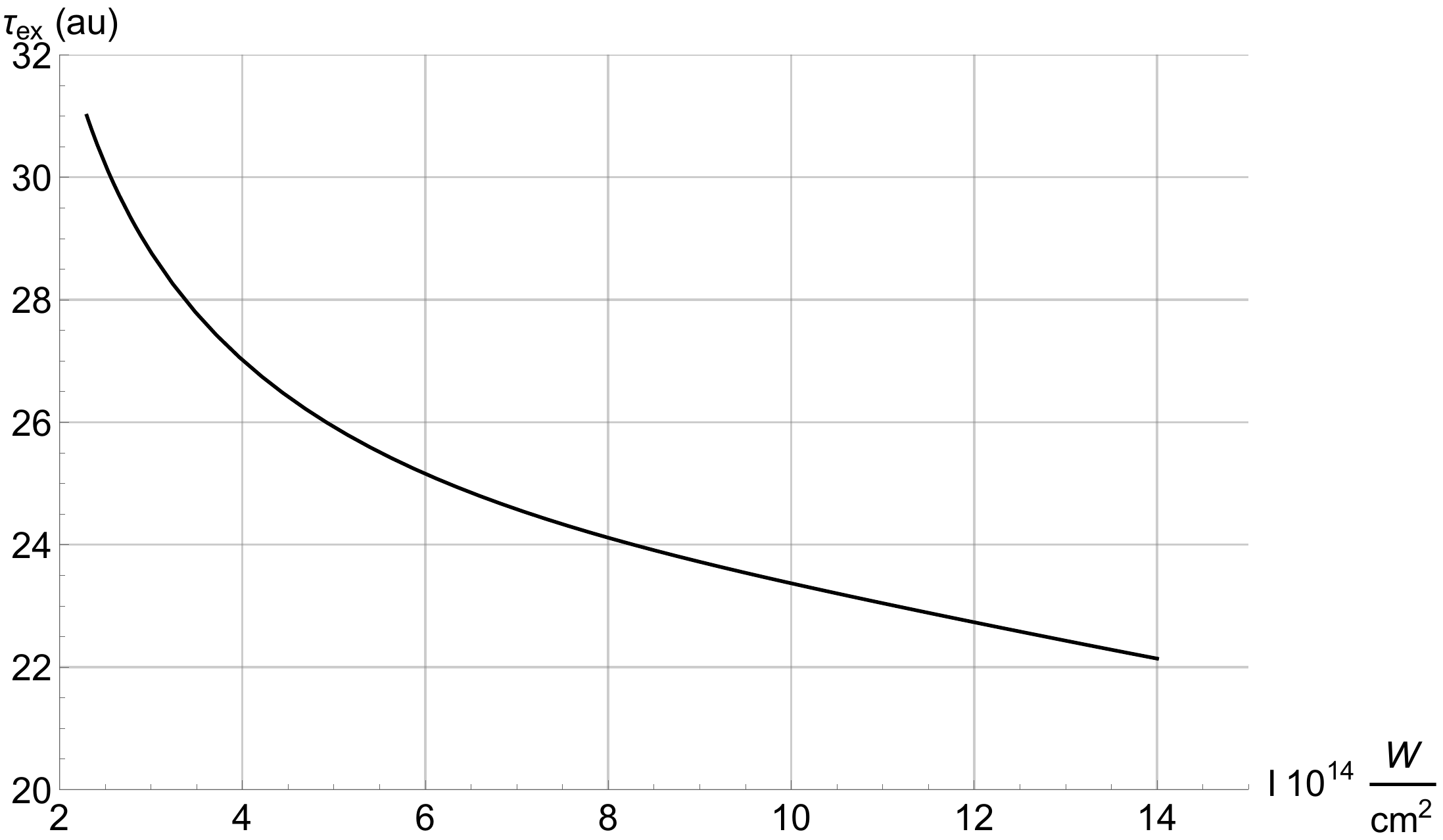}
\caption{Alternative tunnel exit time, based on the fitting process shown in
  Fig.~\ref{fig:trans_size}, as a function of the intensity.}
\label{fig:alttime}
\end{center}
\end{figure}

The transversal fluctuation during the tunneling process is shown in
Fig.~\ref{fig:trans_size}, together with two linear fits of the first and
final stages. The resulting tunneling exit times in Fig.~\ref{fig:alttime} are
less than the time of the peak at $t\approx 27$, so that we obtain negative
tunneling ionization times based on this criterion, similar to
Fig.~\ref{fig:ion_time}. However, the extrapolated time in
Fig.~\ref{fig:trans_size} lies somewhere in the middle of the second stage,
and therefore does not mark the end of the tunneling process.

\begin{figure}
\begin{center} 
\includegraphics[scale = 0.6]{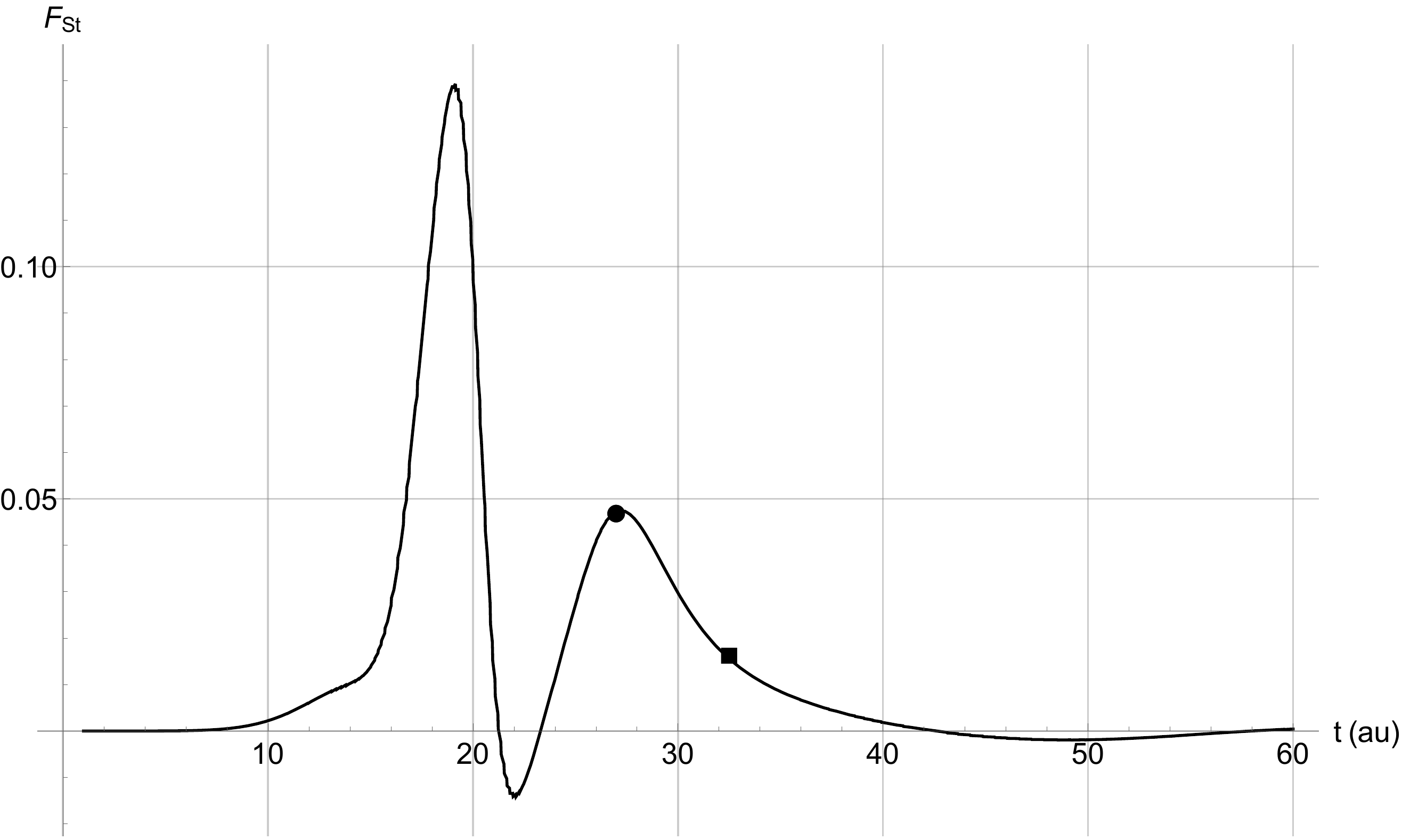}
\caption{The effective force acting on the transverse fluctuations. We see a
  rich structure in the force as the particle goes through the tunneling
  region. The filled circle and square represent the last local maximum and
  the inflection point, respectively. \label{fig:F_trans}} 
                       \end{center}
        \end{figure}

\begin{figure}
\begin{center} 
\includegraphics[scale = 0.8]{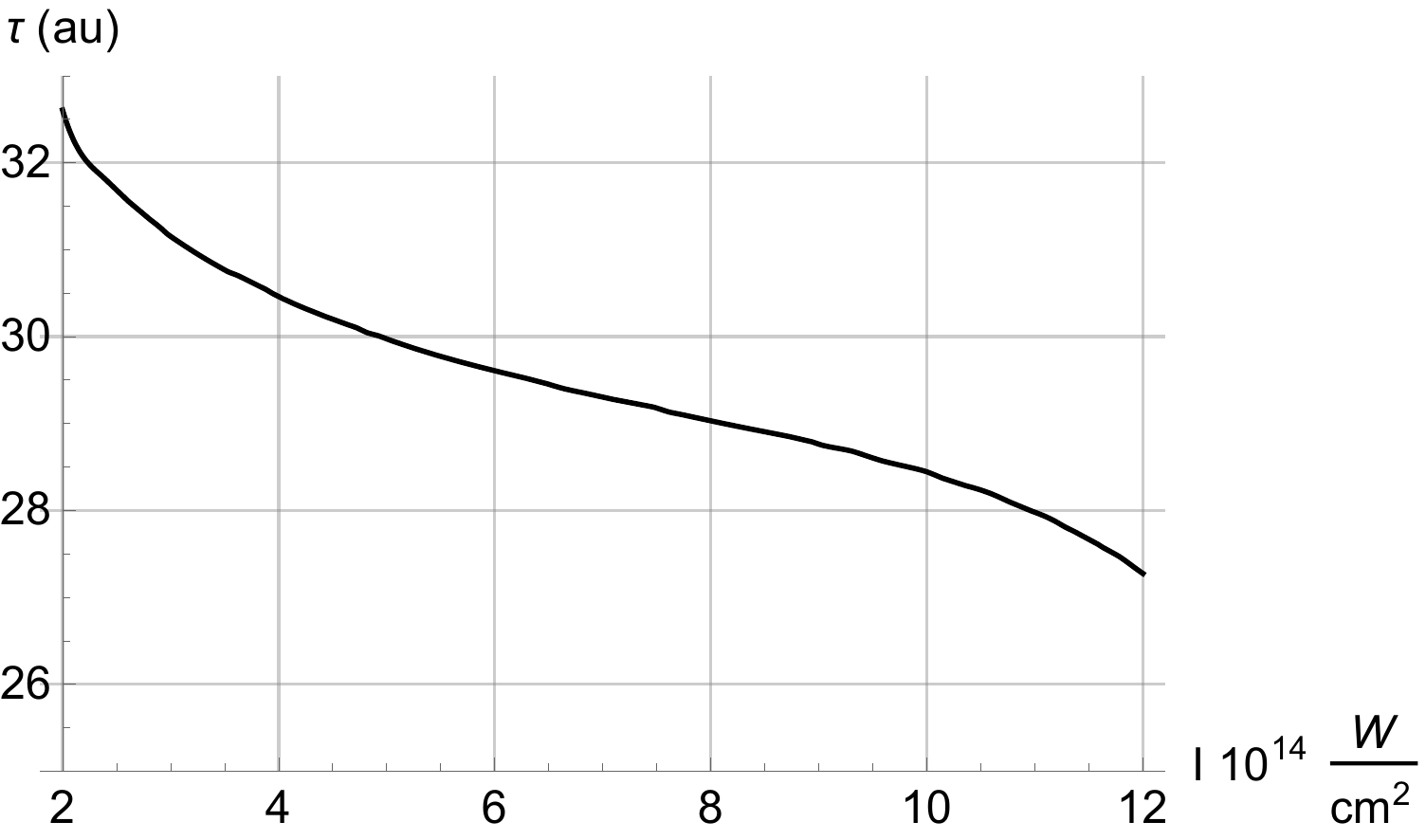}
\caption{Tunneling time based on the last inflection point of the tunneling phase force.} \label{fig:AltAlt_time} 
                       \end{center}
        \end{figure}

We have to look at the tunneling dynamics in more detail in order to identify
the end of tunneling. In Fig.~\ref{fig:F_trans} we show the second time
derivative of the transversal fluctuation as a function of time, which can be
interpreted as an effective force that causes the spreading. The three phases
are clearly visible, with significant time dependence and a rich dynamics only
in the important second phase during which tunneling happens.  The time where
there is a negative force is interesting, because it could be interpreted as a
squeezing the particle state as it passes through the tunnel.  The last local
maximum and the last inflection point, indicated in the plot, are very close
to the wave peak and gives the time of the maximum force on the transverse
fluctuations. In particular, the last inflection point can be used
as an indicator for the tunneling exit. For a range of laser intensities,
the resulting tunneling exit times are shown in Fig.~\ref{fig:AltAlt_time}. In
the entire range shown in this diagram, the exit time is greater than the time
of maximum intensity at $t\approx 27$, and a positive tunneling ionization
time of a few atomic units is obtained.

\section{Summary}

In summary, our main result --- an all-orders effective potential --- makes
possible a detailed analysis of the tunneling dynamics in various
situations. It agrees well with observed features and is able to make new
predictions.  Numerical solutions give us an efficient way of generating data
about the state of the electron which can be compared with observations.  Our
method, perhaps in combination with numerical simulations of multi-electron
wave functions, can therefore be used to turn ionization experiments into
indirect microscopes focused on the atomic state.
 
We have found qualitative agreement between our approximation and the exact
Bohmian treatment. In particular, there is always a tunneling delay. One
advantage of our new methods is that we have a single effective trajectory
describing the quantum state through its expectation values and moments. This
trajectory can directly be compared with the classical back-propagated
trajectory, showing crucial deviations near the tunneling exit. In specific
examples, classical back-propagation tends to underestimate the tunneling exit
time. Our results therefore indicate non-zero tuneling times, but by about an
order of magnitude less than what had initially been extracted from
experiments. In particular, the tunneling time in a half-cycle pulse is
significantly less than the tunneling time in a static field at a level of the
maximum field of the pulse, which is not surprising once the importance of
non-adiabatic effects has been realized \cite{BackProp,Bohmian}.

We also found that the definition of tunneling ionization time in non-constant
fields, given by the difference of the tunneling exit time and the time of
maximal field strength, does not give a full picture of the tunneling
dynamics. In particular, it is possible for the electron to start tunneling
well before the maximum field is reached. The entire tunneling process then
takes longer than indicated by the tunneling ionization time, considered
mainly in \cite{BackProp2}. The tunneling traversal time, used in
\cite{Bohmian}, gives a more complete picture of time-dependent tunneling. In
our examples, we see that a tunnel opens up already at weak fields: The
intensity assumed in the static example of Fig.~\ref{ContourPlot} is about one
tenth of the intensity used in our non-static examples, such as
Fig.~\ref{fig:energy}; see also Fig.~\ref{fig:ContourZoom}. 

Unfortunately, it is difficult to extract the full traversal time from
experiments, but we have given examples of indirect signatures, such as the
spot size based on fluctuations, which could be useful in this
context. Moreover, if the spot size and a corresponding longitudinal
fluctuation can be measured, one could use it, along with the final
expectation values of position and momentum, as initial conditions for {\em
  semiclassical backpropagation} defined as in \cite{BackProp} but using our
effective dynamics instead of the classical dynamics. This process would
eliminate potential problems of classical backpropagation near turning points.

\section*{Acknowledgements}

This work was supported in part by NSF grant PHY-1607414.


\end{document}